\documentclass[12pt]{article}
\usepackage{amssymb,amsmath}
\usepackage{comment}
\usepackage[dvipdfm]{graphicx}
\usepackage{subfigure}
\usepackage{booktabs}
\begin{document}

%%%%%%%%%%%%%%%%%%%%%%%%%%%%%%%%%%%%%%%%%%%%
\thispagestyle{empty}
\begin{flushright}
KEK-TH 1468
\\
KUNS-2351\\
OU-HET 710
%\\
%July 15, 2010
%arXiv:yymm.nnnn
\end{flushright}
\vskip1cm
\begin{center}
{\Large Evidence for Duality of Conifold from
Fundamental String}

\vskip2cm
Takahiro Nishinaka,$^a$\footnote{nishinak[at]post.kek.jp} Takashi Okada,$^b$\footnote{okada[at]gauge.scphys.kyoto-u.ac.jp} Tadashi Okazaki,$^c$\footnote{tadashi[at]het.phys.sci.osaka-u.ac.jp} and Satoshi Yamaguchi$^c$\footnote{yamaguch[at]het.phys.sci.osaka-u.ac.jp}

\bigskip\bigskip
$^a$
{\it High Energy Accelerator Research Organization (KEK), \\
Tsukuba, Ibaraki, 305-0801, Japan}
\\
\vskip2mm
$^b$
{\it Department of Physics, Kyoto university, Kyoto 606-8502, Japan}
\\
\vskip2mm
$^c$
{\it Department of Physics, Graduate School of Science, Osaka University, Toyonaka, Osaka 560-0043, Japan}
\end{center}

%%%%%%%%%%%%%%%%%%%%%%%%%%%%%%%%%%%%%%%%%%%%
\vskip2cm
\begin{abstract}
We study the spectrum of BPS D5-D3-F1 states in type IIB theory, which are proposed to be dual to D4-D2-D0 states on the resolved conifold in type IIA theory. We evaluate the BPS partition functions for all values of the moduli parameter in the type IIB side, and find them completely agree with the results in the type IIA side which was obtained by using Kontsevich-Soibelman's wall-crossing formula. Our result is a quite strong evidence for string dualities on the conifold.
\end{abstract}

%%%%%%%%%%%%%%%%%%%%%%%%%%%%%%%%%%%%%%%%%%

\newpage
\setcounter{tocdepth}{2}
%\tableofcontents
%%%%%%%%%%%%%%%%%%%%%%%%%%

\section{Introduction and summary}
%%D-brane configuration and String%%%%%%%%%%
%\subsection{Introduction}

%Duality
String dualities give us many interesting equivalences between different expressions for string theory.
%%% ies. They help us to analyze non-perturbative regions.
%T-duality of conifold
For example, the T-duality relates type IIA and IIB string theory, where the background fields in both sides are non-trivially related to each other. %%%
The S-duality is a more non-trivial example of the string dualities, which relates the weak and strong coupling regimes of type IIB string theory. %%%
%%%It is well known that the T-duality leads to the existence of D-branes, however we can extract other interesting facts. For example, as shown in \cite{}, T-duality can relate conifold geometries to type IIA brane configurations including NS fivebranes.
 %S-duality of type IIB
%%%On the other hand, S-duality transforms the weak coupling theories into the strong coupling ones.
Since it involves the strong coupling regime in one side, 
%%% Unlike the T-duality,
in order to test the S-duality we need some non-perturbative analysis beyond the %%%usual 
perturbative string theory. 
%%% because it includes strong couplings.

%BPS-state wall crossing phenomena
One of the promising way to test the S-duality is to examine BPS states.
In general, BPS states are important probes of non-perturbative properties of the theories with extended ($\mathcal{N}\ge 2$) supersymmetry.
%The reason for this is that the BPS states belong to a short multiplet of supersymmetry and quantum corrections to their physics are highly constrained. 
%%%Their mass is determined by their charges as a consequence of the SUSY algebra.
%%%Although the 
%As a result, t
The BPS states are generically stable
under the change of the hyper multiplet moduli parameters, such as the string coupling constant.
%because of the energy and charge conservations.
However, their degeneracy
%%%of BPS states
can actually ``jump'' at some codimension one subspace in the vector multiplet moduli space. 
%%% as we change the value of K\"{a}hler moduli.
 This is called wall-crossing phenomena, which make it valuable and interesting to study counting the BPS states.
The BPS counting problem and the wall-crossing phenomena are important also in many other research
 areas, such as black hole microstates \cite{Denef:2000nb, Denef:2007vg, deBoer:2008fk, deBoer:2008zn, Cardoso:2008ej, VanHerck:2009ww, Andriyash:2010qv, Andriyash:2010yf, Manschot:2010qz, Manschot:2011xc}, Donaldson-Thomas invariants \cite{Szendroi:2007nu, Nagao:2010kx, Nagao, Collinucci:2008ht, Jafferis:2008uf, Chuang:2008aw, Kontsevich-Soibelman, Ooguri:2008yb, Dimofte:2009bv, Chuang:2009pd, Kontsevich:2009xt, Sulkowski:2009rw, Dimofte:2009tm, Krefl:2010sz, Aganagic:2010qr}, topological strings and instanton counting \cite{Collinucci:2009nv, Nagao:2009rq, Cecotti:2009uf, Szabo:2009vw, Chuang:2010wx, Bringmann:2010sd, Manschot:2010nc}, M-theory viewpoint \cite{Aganagic:2009kf, Aganagic:2009cg}, exact counting of $\mathcal{N}=4$ dyons \cite{Cheng:2007ch, Sen:2008ht, Cheng:2008fc, Cheng:2008kt, Cheng:2009hm}, supersymmetric gauge theories \cite{Gaiotto:2008cd, Gaiotto:2009hg, Gaiotto:2010be, Gaiotto:2011tf} and many others \cite{Diaconescu:2007bf, Jafferis:2007ti, Andriyash:2008it,  David:2009ru, Manschot:2009ia, Manschot:2010xp, Chuang:2010ii, Nishinaka:2010qk, Nishinaka:2010fh, Alim:2010cf, Nishinaka:2011sv}.

In this paper, we test the string dualities by studying the BPS states of the theory. In particular, we compare the spectra of BPS states in both sides of the duality. In one side, we consider the D4-D2-D0 system in type IIA string theory on the resolved conifold. After taking the T and S-duality, this D-brane system is mapped to a D5-D3-F1 system in type IIB theory on $\mathbb{R}^{1,8}\times S^1$. The original conifold geometry is mapped to two NS5-branes after the T-duality transformation \cite{Uranga:1998vf,Bershadsky:1995sp} and then the S-duality changes them into two D5-branes. Our purpose in this paper is to check the equivalence of the BPS spectra in both sides of the duality, including the effects of the wall-crossing phenomena. %%%

%K&S formula and d4-d2-d0 1
%In \cite{}, Kontsevich and Soibelman proposed the mathematical wall crossing formula (KS formula) which tell us how we can determine the degeneracy of BPS states as we change the moduli.
We start from the type IIA side, where 
%%%Namely BPS partition function BPS bound states of one non-compact D4-brane and arbitrary numbers of D2 and D0-branes on a Calabi-Yau three-fold $X$ were studied and their partition function were given.
our D-brane bound states of interest are composed of one non-compact D4-brane and various numbers of D2 and D0-branes bound to it. We let our D2-branes be wrapped on the compact two-cycle of the conifold so that the D2-branes have finite mass, while the D4-brane is stretched in a non-compact divisor of the conifold. The BPS partition function is defined by
\begin{equation}
\mathcal{Z}_{\rm BPS}(u,v):=\sum_{Q_{0}, Q_{2}\in \mathbb{Z}}\Omega(\mathcal{D}+Q_2\beta-Q_0dV)u^{Q_{0}}v^{Q_{2}},\label{eq:partition_function}
\end{equation}
where $\Omega(\gamma)$ denotes the BPS index of charge $\gamma$ and $\mathcal{D}\in H^{2}(X)$ stands for one unit of the non-compact D4-brane charge. 
The unit D2 and D0-brane charges are denoted by $\beta \in H^{4}(X)$ and $-dV\in H^{6}(X)$ respectively. Therefore $Q_{2}, Q_{0} \in \mathbb{Z}$ are D2 and D0-brane charges of the bound states. The four-form $\beta$ is dual to the compact two-cycle of the conifold. 
The Boltzmann weights for D2 and D0 branes are denoted by $v$ and $u$ respectively.
%K&S formula and d4-d2-d0 2

This BPS partition function of the D4-D2-D0 states was already evaluated in \cite{Nishinaka:2010qk} by using the Kontsevich-Soibelman's wall-crossing formula \cite{Kontsevich-Soibelman, Kontsevich:2009xt}.
%\footnote{The KS-formula tells us how the index of the BPS states jumps as we change the vacuum moduli.}
%%%In the case of resolved conifold, corresponding to K\"{a}her moduli $t=\infty$, we can determine the BPS partition function 
As was previously mentioned, the partition function $\mathcal{Z}_{\rm BPS}$ depends on the moduli parameter $z$, whose imaginary part ${\rm Im}\,z$ is roughly the size of the compact two-cycle of the conifold.\footnote{To be more precise, we set $z$ so that the central charge of a D2-brane is equal to $z$. When ${\rm Im}\,z$ is large, ${\rm Re}\,z$ and ${\rm Im}\,z$ are regarded as the B-field and the size of the compact two-cycle of the conifold.} Suppose we fix ${\rm Re}\,z = 1/2$ and move ${\rm Im}\,z$ from ${\rm Im}\,z = +\infty$ to ${\rm Im}\,z=-\infty$. When the moduli cross the walls of marginal stability, the partition function jumps. For ${\rm Im}\,z>0$, it is given by \footnote{In \cite{Nishinaka:2010qk}, $\prod_{k=1}^\infty(1-u^k)^{-1}$ in equation \eqref{bps1} is replaced by $\prod_{k=1}^\infty(1-u^k)^{1-\chi(C_4)}$ where $\chi(C_4)$ is the Euler characteristics of the four-cycle $C_4$ wrapped by the D4-brane. Since our D4-brane is non-compact, the Euler characteristics $\chi(C_4)$ generally has an ambiguity. However, in this paper, we omit all such ambiguities coming from the non-compactness and simply set $\chi(C_4) = \chi(\mathbb{P}^1) = 2$.} 
%\footnote{From the analysis of \cite{},
%when putting the non-compact D4-brane on a divisor $C_{4}$ and making the compact two-cycle of the conifold shrink to zero size,
% we obtain
%\[
%Z_{BPS}(u,v)=\prod_{n=1}^{\infty}\left(
%\frac{1}{1-u^{n}}\right)^{\chi(C_{4})-1}\prod_{m=0}^{\infty}(1-u^{m}v),
%\]
%where $\chi(C_{4})$ is the Euler characteristic of the divisor $C_{4}=(\textrm{total space of }\mathcal{O}(-1)\rightarrow \bm{P}^{1})$.
%Since the divisor is non-compact, its Euler characteristic cannot be fixed. 
%However we here simply set $\chi(C_{4})=\chi(\bm{P}^{1})=2$ by neglecting all the ambiguities.}
\begin{equation}
\mathcal{Z}_{\rm BPS}(u,v) = \prod_{k=1}^\infty\left(\frac{1}{1-u^k}\right)\prod_{m=0}^{\infty}(1-u^{m}v)\prod_{n=1}^{n_0}(1-u^{n}v^{-1}).
\label{bps1}
\end{equation}
The integer $n_0>0$ denotes the number of wall-crossings that occur when we move the moduli from ${\rm Im}\,z = 0$ to its given value.
On the other hand, for ${\rm Im}\,z \leq 0$, the partition function can be written as
\begin{equation}
\mathcal{Z}_{\rm BPS}(u,v) = \prod_{k=1}^\infty\left(\frac{1}{1-u^k}\right)\prod_{m=m_0}^{\infty}(1-u^{m}v),
\label{bps2}
\end{equation}
where $m_0$ is the number of walls between the given value of ${\rm Im}\,z (\leq 0)$ and ${\rm Im}\,z=0$.
%\begin{equation}
%Z_{BPS}=f(u)(1-v)\prod_{m=1}^{\infty}(1-u^{m}v)(1-u^{m}v^{-1}).
%\label{bps1}
%\end{equation}
%By using the KS-formula, in the case of $t=0$ we have
%\begin{equation}
%Z_{BPS}=f(u)\prod_{m=1}^{\infty}(1-u^{m}v),
%\end{equation}
%and in the case of $t=-\infty$ we have
%\begin{equation}
%Z_{BPS}=f(u).
%\end{equation}

Now, our remaining task is to evaluate the BPS partition function in the dual type IIB side, and compare the result with the above expressions in the type IIA side. This is the main subject of this paper. 

%\subsection{Conclusion}
%Our works
%In this paper we prove S duality by using BPS counting analysis. As a result, we can get beautiful stringy picture of wall crossing phenomena.
%Firstly we start from conifold system with D4-D2-D0 wrapped.
%Secondly we transform it into D3-D5-D5' system in flat space-time 
%by using S and T duality.

%%%In this paper, 
Let us here summarize our results in this paper. We study the spectrum of the BPS D5-D3-F1 states in type IIB theory, which are dual to the above D4-D2-D0 states on the conifold.
% which is dual to the D4-D2-D0 system analyzed in \cite{Nishinaka:2010qk}
%, and found that the BPS spectrum is exactly the same as \eqref{bps1}
%and \eqref{bps2} including wall crossing. 
One of the most important advantages of the type IIB side is that the BPS spectrum can be analyzed in the perturbative open string theory on the flat spacetime $\mathbb{R}^{1,8}\times S^1$. The D0 and D2-brane charges in the type IIA side are identified with the winding number of the fundamental strings and the electric charge on the D5-brane, respectively.
%Then by noting that the dualized D3-D5-D5'-branes can be analyzed by our familiar string theory, type IIB superstring theory,
We examine the spectrum of the BPS fundamental strings stretched between D3 and D5-branes, and
%%%From this we 
calculate the BPS partition function. The result is exactly the same as (\ref{bps1}) and \eqref{bps2}, including the wall-crossing phenomena.  This result is a quite strong evidence of the duality.
% which implies that the BPS spectra are equivalent in both sides of the duality, including the effects of the wall-crossing phenomena.
%%% in the case of resolved conifold. 
%%% Finally we add electric field, which is identified with B-field in original theory, in dualized D3-D5-D5' system.

We also show that the moduli parameters in the type IIA side can be identified with the relative positions of the D5-branes and the electric field on the D3-brane in the type IIB side.
This gives us the pictorial understanding of the wall-crossing phenomena of the original D4-D2-D0 system.
%%%in string theory.
%%%because
%%%In particular, turning on
%%%added
%%%the electric field on the D3-brane has the effect of changing the angles of open strings which end on the D3-brane.
%%%Our analysis can be regarded as an evidence for the two validness for 
%%%\begin{enumerate}
%%%\item the duality of the conifold,
%%%\item Kontsevich Soibelman formula.
%%%\end{enumerate}

%For future work, .............

%construction

The construction of this paper is as follows.
In Section \ref{sec:duality}, we discuss the dualizing procedure of our conifold system with D4-D2-D0 branes.
%%%Because of this discussion,
%The relevant BPS states can be regarded as the perturbative string states.
We identify the relevant moduli and the charges in the dual picture.
Section \ref{sec:open-string} and section \ref{sec:wall-crossing} involve the main result of this paper. 
We count BPS states by investigating BPS open string spectrum on the D5-D3 system and obtain the BPS partition functions in all the chambers.  They completely agree with the BPS spectrum in the type IIA side.
%The wall-crossing phenomena can be seen pic
 %As a consequence, we can justify the duality of conifold. 
%Also by adding the electric field, we obtain the beautiful picture of wall-crossing phenomena in string theory.
%In Section4 we discuss our results and give a few comments.

\section{Sequence of dualities}
\label{sec:duality}
In this section, we consider the type IIB dual of the D4-D2-D0 states on the conifold, which will turn out to be D5-D3-F1 states. In the original type IIA side, we put one D4-brane on a non-compact holomorphic four-cycle $C_4$ of the conifold $\mathcal{O}(-1)\oplus\mathcal{O}(-1) \to \mathbb{P}^1$, and consider D2-branes wrapped on the rigid $\mathbb{P}^1$ and D0-branes localized in the conifold. These D-brane bound states can be seen as charged particles in four-dimensional spacetime transverse to the six-dimensional conifold geometry. Their BPS partition function was evaluated in \cite{Nishinaka:2010qk} as \eqref{bps1} and \eqref{bps2}. 

Below, we show that this brane configuration is dual to a D5-D3-F1 system after the T and S-dual transformations. We also discuss how the moduli parameters of the theory are mapped under the duality transformations.

%%%In section 3, we evaluate the spectrum of these BPS bound states in the dual type IIB side. We will see that the calculation in the type IIB side can be performed in the perturbative open string theory.

\subsection{Configuration of D-branes and open strings}

Our staring point is the D4-D2-D0 bound states
%%%following brane configuration
 in type IIA string theory on the conifold. The brane configuration is summarized in Table \ref{table:D4-D2-D0}.
%%%and we will study the BPS bound states of a  D4-brane and D2 and D0 branes on the conifold by applying a sequence of dualities. 
\begin{table}
\begin{center}
\begin{tabular}{|c||c|c|c|c||c|c|c|c|c|c|}\hline
&$x_0$&$x_1$&$x_2$&$x_3$& \multicolumn{6}{|c|}{the conifold} \\ \hline \hline
D4&$\circ$& &  & &\multicolumn{6}{|c|}{wrapping on $C_4$
%%%the non-compact 4-cycle
} \\ \hline 
D2&$\circ$& & & &\multicolumn{6}{|c|}{wrapping on $\mathbb{P}^1$
%%%the compact 2-cycles
}\\ \hline
D0&$\circ$& &  & &\multicolumn{6}{|c|}{} \\ \hline
\end{tabular}
\caption{The BPS D-brane configuration of interest in the original type IIA setup. There is one D4-brane wrapped on the non-compact divisor $C_4$ of the conifold. The D2-branes are wrapped on the rigid $\mathbb{P}^1$ while the D0-branes are localized in the conifold.}
\label{table:D4-D2-D0}
\end{center}
\end{table}
Here $C_4$ denotes a non-compact four-cycle of the resolved conifold, whose topology depends on the moduli parameter $z$. In fact, $C_4$ is topologically equivalent to $\mathcal{O}(-1)\to\mathbb{P}^1$ and $\mathbb{C}^2$ in the moduli regions ${\rm Im}\,z > 0$ and ${\rm Im}\,z <0$, respectively. This is related to the flop transition of the conifold, and its relation to the wall-crossing phenomena was discussed in \cite{Nishinaka:2010qk}. 

Now, we
%%%first
take the T-duality transformation along a $U(1)$ orbit of the conifold.
%%%\cite{uranga1999brane}.
Recall that the defining equation of the conifold is
\begin{equation}
xy=zw,
\end{equation}
which can be interpreted as a 
%%%$\IC^*$
$\mathbb{C}^*$-fibration over
%%%$\IC^2$
$\mathbb{C}^2$
%%%represented
parameterized by $z,\ w$. That is, for given $z,\ w$, we can relate 
$x$ and $y$ via this equation.
There is a $U(1)$ orbit in the
%%%$\IC^*$
$\mathbb{C}^*$ fiber, which is generated by 
\begin{equation}
x \rightarrow \lambda x,\ y\rightarrow \lambda^{-1} y, \label{eq:C-star}
\end{equation}
where $\lambda$ is a complex number with $|\lambda|=1$. 
We take
%%%By taking
the T-duality transformation along this orbit, which leads to two NS5-branes in flat spacetime \cite{Uranga:1998vf}.
%%%the degeneration in the fibration is mapped into  two NS5-branes in flat spacetime, denoted by NS5 and NS5', and D2 and D0-branes are replaced by D3 and D1-respectively.
The original D2-branes become D1-branes ending on the NS5-branes, while
D0-branes change into winding D1-branes in the type IIB side. On the
other hand, the D4-brane wrapped on $C_4$ becomes D3-brane ending on one of the NS5-branes.
The brane configuration after the T-dual transformation is summarized in Table \ref{table:NS5-D3-D1}.
%%%then becomes
\begin{table}
\begin{center}
\begin{tabular}{|c||c|c|c|c||c|c|c|c|c|c|}\hline
&$x_0$&$x_1$&$x_2$&$x_3$&$x_4$&$x_5$&$x_6$&$x_7$&$x_8$&$x_9 (S^1)$ \\ \hline \hline
NS5&$\circ$ &$\circ$&$\circ$&$\circ$ & $\circ$ &$\circ$ & & & &\  \\ \hline
NS5'&$\circ$&$\circ$&$\circ$&$\circ$ & & &$\circ$ &$\circ$ & &\  \\ \hline
D3 &$\circ$& & & &  &  & $\circ$ & $\circ$ & $\circ$ &\  \\ \hline
D1 &$\circ$& & & & & & & & &$\circ$  \\ \hline
\end{tabular}
\caption{The brane configuration after the T-dual transformation. The original conifold geometry is mapped to two NS5-branes, while the D4-brane becomes a D3-brane. The original D2 and D0-branes change into D1-branes.}
\label{table:NS5-D3-D1}
\end{center}
\end{table}
%%%Note that we have taken the directions which the two NS5-branes span by 012345 and 012367.
Here $x^9$-direction is compactified on a circle for the T-duality transformation.

\begin{figure}
%%%[htbp]
\begin{center}
\includegraphics[width=8cm]{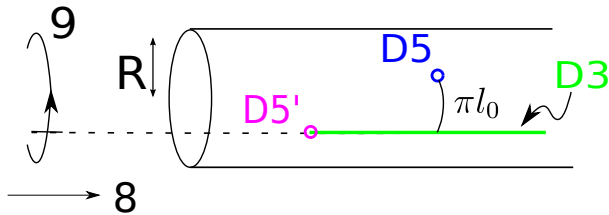}
\caption{Dualized configuration of D3-D5-D5'-branes. Green line is D-3 brane. Blue and pink points are D5 and D5'-branes respectively. The radius of $S^{1}$ in 9 direction is $R$ and the distance between D3 and D5 in 9 direction is $\pi l_{0}$. }
\label{fig:setup}
\end{center}
\end{figure}

We further perform the S-duality transformation and obtain a D5-D3-F1 system as in Table \ref{table:D5-D3-F1}.
%%%Further, by performing the S-duality, the NS5-branes are replaced by two D5-branes, and D1-branes by fundamental strings.
%%%We thus obtain the following configuration(see Fig.\ref{fig:setup} ). 
\begin{table}
\begin{center}
\begin{tabular}{|c||c|c|c|c||c|c|c|c|c|c|}\hline
&$x_0$&$x_1$&$x_2$&$x_3$&$x_4$&$x_5$&$x_6$&$x_7$&$x_8$&$x_9(S^1)$  \\ \hline \hline
D5&$\circ$&$\circ$&$\circ$&$\circ$& $\circ$ &$\circ$& & & & 	 \\ \hline
D5'&$\circ$&$\circ$&$\circ$&$\circ$& & &$\circ$& $\circ$	&\  	 &\  	\\ \hline
D3&$\circ$& & & & & &$\circ$ &$\circ$ &$\circ$ &  	\\ \hline
F1&	$\circ$& & & & 	&  & & & &$\circ$ \\ \hline
\end{tabular}
\caption{The brane configuration after the T and S-dual transformations. Now, the whole system only involves D-branes and fundamental strings, which can be analyzed in the perturbative open string theory.}
\label{table:D5-D3-F1}
\end{center}
\end{table}
This brane configuration can be depicted as in Fig.~\ref{fig:setup}.
The locations of two D5-branes correspond to two degenerating points of the $\mathbb{C}^*$-fibration of the conifold, in type IIA language. The D3-brane ends on a D5-brane denoted by D5' because the original D4-brane is wrapped on the divisor $\mathcal{O}(-1)\to\mathbb{P}^1$ in the conifold.\footnote{When ${\rm Im}\,z<0$, the divisor $C_4$ wrapped by the D4-brane is topologically $\mathbb{C}^2$. For the moduli dependence of this brane configuration, see subsection \ref{subsec:moduli}.} Our D3-brane is extended to $x^8=+\infty$ as in figure \ref{fig:setup}. The fundamental strings are stretched between the D5 and D3-brane. Here, the original D0-brane charge is identified with the winding number of the fundamental strings along $x^9$-direction, while the D2-brane charge is mapped to the electric charge on the D5-brane. 

\begin{figure}
%%%[htbp]
\begin{center}
\subfigure[3-5 string]{\includegraphics*[width=.3\linewidth]{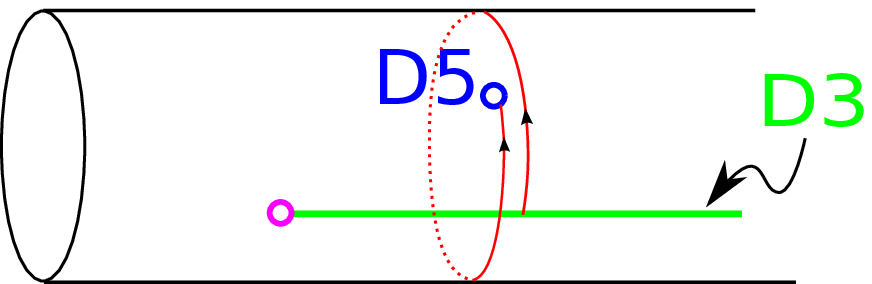}
\label{fig:35}}
\subfigure[5-3 string]{\includegraphics*[width=.3\linewidth]{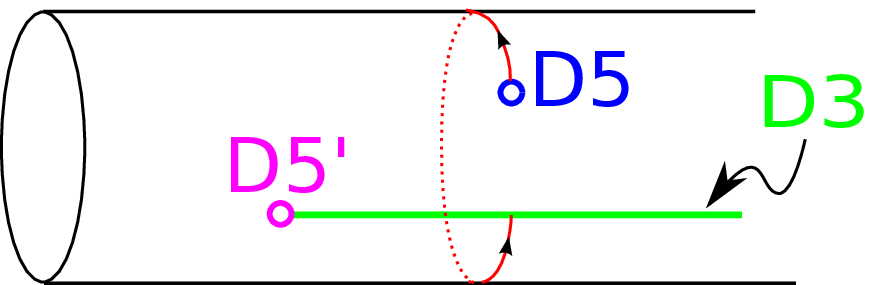}
\label{fig:53}}
\subfigure[3-3 string]{\includegraphics*[width=.3\linewidth]{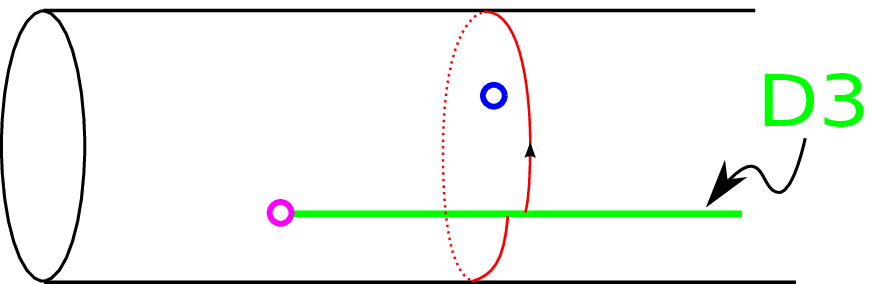}
\label{fig:33}}
\end{center}
\caption{Three kinds of open strings relevant for the BPS counting in the type IIB side. The original D2 and D0-brane charges are now mapped to the electric charge on the D5-brane and the winding number of the strings, respectively.
%%%To be precisely, this is a sketch in the absence of an electric field background $F_{08}$. As is discussed later, a non-zero $F_{08}$ tilts the open string in $x^8$-$x^9$ plane.
}
\label{fig:three}
\end{figure}

Note that this configuration only involves D-branes and fundamental strings in flat spacetime which can be analyzed in the perturbative open string theory. 
In section 3, we will
%%%analyze the BPS spectra of D4-D2-D0 bound states.
evaluate the BPS partition function for this brane configuration, by counting the number of stable BPS strings on the D5-D3 system.
There are three kinds of relevant open strings, that is, D3-D5, D5-D3 and D3-D3 strings (see Fig.~\ref{fig:three}).

Let us briefly mention the BPS states without D3-brane. This corresponds to BPS D2-D0 bound states in the type IIA side. There are two types of the BPS states: open strings connecting two D5-branes (D5-D5' string) and closed winding strings localized on either D5 or D5'-brane. The D5-D5' string with a winding number becomes a half hyper multiplet in four dimensions and contributes $1$ to the BPS index. This number is consistent with the result $\Omega(\pm\beta-n dV)=1,\ n\in \mathbb{Z}$ in the type IIA side (see for example \cite{Jafferis:2008uf}).  On the other hand, the existence of the winding closed strings is consistent with the BPS indices $\Omega(-ndV)=-2,\ n\in \mathbb{Z}$ in the type IIA side.\footnote{This counting is actually ambiguous since the internal space is non-compact here.}

\subsection{Moduli parameters}
\label{subsec:moduli}

We here briefly discuss the moduli parameters of the theory. Since our BPS bound states can be seen as charged particles in four dimensions spanned by $x^0,x^1,x^2$ and $x^3$, the spectrum of the BPS states depends on the vector multiplet moduli of the $d=4, \mathcal{N}=2$ supersymmetric theory. In the original type IIA setup, this vector multiplet moduli correspond to the K\"ahler moduli of the conifold, which we denote by $z$. For large ${\rm Im}\,z$, the real and imaginary parts of $z$ are the B-field and size of the rigid $\mathbb{P}^1$ of the conifold, respectively. 

In order to interpret this moduli $z$ in the type IIB side, let us consider a D2-brane wrapped on the rigid $\mathbb{P}^1$ of the conifold. Such a D2-brane is mapped to a fundamental string stretched between two D5-branes after the T and S-duality transformations. Since it originally has no D0-brane charge, the dual fundamental string has the vanishing winding number along $x^9$-direction. The mass of this fundamental string is, of course, equal to the distance between two D5-branes,\footnote{We can show that there is no non-vanishing B-fields in $x^8$-$x^9$-plane in the type IIB side, after taking the S-duality transformation.}
while the mass of the original D2-brane is proportional to $|z|$. So, we find that the distance between two D5-branes in Fig \ref{fig:setup} can be written as $|z|$ up to a prefactor. To be more precise, if we set $x^8$ and $x^9$ so that the D5' is located at the origin on the $x^8$-$x^9$ plane, the position of the D5 is determined by
\begin{align}
x^9  + ix^8 =2\pi R z,\label{eq:D5-position}
\end{align}
where $R$ is the radius of $S^1$ on which $x^9$-direction is compactified. Note here that the periodicity with respect to the B-field
\begin{eqnarray}
{\rm Re}\,z \sim {\rm Re}\,z + 1
\end{eqnarray}
can be understood as a periodicity in the compactified direction $x^9$. Thus, the moduli parameter $z$ is interpreted as the position of the D5-brane in $x^8$-$x^9$ plane.\footnote{We should here set $x^8 = +2\pi R\, {\rm Im}\,z$ rather than $x^8=-2\pi R\, {\rm Im}\,z$, because we have assumed that the D3-brane is extended to $x^8=+\infty$ (See Fig.~\ref{fig:setup}). Recall that for ${\rm Im}\,z>0$ our D4-brane is wrapped on $\mathcal{O}(-1)\to \mathbb{P}^1$, and therefore two degenerating points of the $U(1)$-fibration of the conifold are embedded in the divisor wrapped by the D4-brane. This implies that, in the type IIB side, both the D5 and D5'-branes should have the non-vanishing intersection to the D3-brane if ${\rm Re}\,z=0$ and ${\rm Im}\,z>0$.} 

%%%The BPS spectra in general depend on the moduli parameters. The relevant moduli for wall-crossing phenomena is the K\"ahler moduli, and here we write the moduli parameters of this system and explain how they transform under the dualities. 

%%%Since the resolved conifold has only one compact 2-cycle, we have one K\"ahler moduli $z=B+iJ$ for the compact cycle, where $B$ is the $B$-filed on the 2-cycle and $J$ is its size. For the non-compact cycles, on the other hand, we also have another moduli associated with the phase of the K\"ahler moduli for the non-compact cycles.

One subtlety here is that there is another type of ``modulus,'' which is
a remnant of the K\"ahler parameters for non-compact cycles of the
conifold. In particular, we can turn on the B-field $B\not\in\mathbb{Z}$
on the non-compact four-cycle $C_4$ wrapped by the D4-brane. We choose
$B$ to be transverse to the B-field for the rigid $\mathbb{P}^1$, namely
${\rm Re}\,z$, and fix it to be non-vanishing as in
\cite{Nishinaka:2010qk}.\footnote{This new ``modulus'' can be understood
by considering the conifold as a local limit of some compact Calabi-Yau
manifold.  If we collectively write the K\"ahler moduli of the
(would-be) non-compact cycles as $\Lambda e^{\i \varphi}$, only
$\varphi$ remains as a ``modulus'' after taking the local limit $\Lambda
\rightarrow \infty$. This extension of the K\"ahler moduli of the local
Calabi-Yau was proposed in \cite{Jafferis:2008uf}. Note that our
non-vanishing B-field for the non-compact cycles implies that $\varphi
\not\in \pi\mathbb{Z}$, which is the same condition imposed in
\cite{Nishinaka:2010qk}.} The counterpart of this B-field in the type
IIB side can be identified as follows. Since it is transverse to the
rigid $\mathbb{P}^1$, the T-duality transformation along the
$\mathbb{C}^*$-orbit \eqref{eq:C-star} leaves $B$ invariant. After the
T-duality transformation, $B$ can be regarded as a B-field on the
D3-brane, whose non-vanishing component is $B_{67} = - B_{76}$. There is
a gauge symmetry for the $B$ field and the world volume gauge
field $A$ expressed by $B\to B+d\Lambda,\ A\to A-\Lambda/(2\pi\alpha')$
with a 1-form parameter $\Lambda$. By this gauge transformation, this
B-field $B_{67}$ is transformed into a ``magnetic'' gauge flux $F_{67}$
on the D3-brane.\footnote{This is possible because $x_{6}, x_{7}$
directions are non-compact and thus the magnetic flux is not quantized.}
\footnote{This transformation introduce D1-brane charge (or F-string charge after S-duality) on the D3-brane. However this is irrelevant because we are not counting this D1 or F-string charge.} Then, the S-duality transformation maps this ``magnetic'' flux to the ``electric'' flux $F_{08}$. Hence, the original B-field for the non-compact cycle $C_4$ is mapped to the electric flux $F_{08}$ by the T and S-duality transformations.

%%%After taking the T-duality  along the U(1)-fiber of the resolved conifold, according to Buscher rules\cite{}, the  K\"ahler modulus $z$ is mapped to the position of the two NS5-branes on the cylinder spanned by $x_8$ and $x_9$, and the D3-brane ends one of the two NS5-branes, which we denote by NS5'.\footnote{For the similar analysis, see, for example [,,,,].} Further taking the S-duality, $z$ can be mapped to the position of the two D5-brane on the cylinder. 

 %%%Thus,
In summary, we have one complex moduli parameter $z$ and one real moduli parameter $B$ in the original type IIA setup, which are dual to the position of D5 and the ``electric'' field on the D3-brane, respectively (see Table \ref{table:moduli}).
%%%The moduli $\phi$ is mapped to the B-field on the D3-brane, and this is equivalent to a electric field $F_{08}$ from the world-sheet view point.
\begin{table}
\begin{center}
\begin{tabular}{|c|c|}\hline
Before the TS-duality & After the TS-duality \\ \hline \hline
 K\"ahler moduli $z$ for the rigid $\mathbb{P}^1$ &  position of D5 \\ \hline
%%%the phase associated with the non-compact cycles
 B-field $B$ on the non-compact divisor $C_4$ & ``electric'' field $F_{08}$ on the D3-brane\\ \hline
\end{tabular}
\caption{The vector multiplet moduli parameters in both sides of the TS-duality. }
\label{table:moduli}
\end{center}
\end{table}

\section{Open string spectrum% and Wall-Crossing Phenomena
}
\label{sec:open-string}
In this section, we evaluate the spectrum of our BPS bound states with two D5-branes, a single D3-brane, and fundamental strings (Table \ref{table:D5-D3-F1}). We first note that the configuration of Table \ref{table:D5-D3-F1} always saturates the BPS bound if there is no fundamental string. This is obvious because such a configuration is dual to a single D4-brane wrapped on a holomorphic cycle in the type IIA side. Putting fundamental strings on the D5-D3 system corresponds to adding D2 and D0-brane charges in the original type IIA setup. Thus, the BPS spectrum of interest can be evaluated by counting the number of BPS string states on the D5-D3 system.

%%%In general, it is hard to analyze the string spectrum in a curved background.
%%%However, with help of the TS-duality, we have replaced the D4-D2-D0 system on the conifold by the D3-D5-D5' system in the $flat$ spacetime, and thus can study the string states.
One advantage of our analysis in the dual IIB side is that we only have to analyze the perturbative string theory in {\em flat} spacetime, while analysis in the type IIA side should take into account non-trivial $\alpha'$-corrections due to the curved background.\footnote{Due to the $\alpha'$-corrections, it is difficult to evaluate the BPS spectrum directly in the type IIA side. In \cite{Nishinaka:2010qk}, the authors instead used the Kontsevich-Soibelman formula that tells us how the BPS index changes in wall-crossings.}  It is for this reason that we can explicitly count the number of BPS states for arbitrary values of moduli parameters.

We first note that fundamental strings which we should count is 
%%%there are two types of massive open strings on the D5-D3 system, that is, 
only those on the D3-brane and those stretched between D5 and D3-branes.
%\footnote{Open strings on the D5-brane are not bound to the D3-brane, and therefore we will not count them. What we want to count is BPS fundamental strings bound to D3-branes.} 
This is because we are now considering {\em the bound states} of the D3-brane and
some number of fundamental strings; the fundamental strings must be open strings and at least one end of each string must be attached to the D3-brane.
The first one will be studied in \ref{subsec:D3-D3}, while in \ref{subsec:D5-D3} we analyze the second one.

\subsection{Open strings between D5 and D3-branes}
\label{subsec:D5-D3}

By definition, BPS saturated states lie in the bottom of the mass spectrum with fixed charges. Thus, our BPS string states can be identified with the lowest energy states with the winding number along $x^9$-direction and electric charge on the D5-brane fixed. Recall here that the zero point energies of string states in NS and Ramond sectors can be written as
\begin{eqnarray}
 a_{\rm NS} &=& \left(-\frac{1}{48}-\frac{1}{24}\right)(8-\nu) + \left(\frac{1}{48}+\frac{1}{24}\right)\nu \;\,=\,\; \frac{\nu}{16} - \frac{8-\nu}{16},\label{eq:NS_zero}
\\[2mm]
\qquad a_{\rm R} &=& 0,\label{eq:Ramond_zero}
\end{eqnarray}
respectively, where $\nu$ denotes the number of dimensions in which one of the open string endpoints has the Dirichlet boundary condition and the other has the Neumann boundary condition. 
%This implies that it depends on $\nu$ whether the lowest energy states belong to the NS or R-sector.

We start from strings stretched between the D5 and D3-branes (see Fig.~\ref{fig:three}). For simplicity, we first consider the case of $B=F_{08}=0$ for the electric field on the D3-brane. Recall first that the position of the D5-brane is given by \eqref{eq:D5-position} when the D5'-brane is located at $x^8=x^9=0$. Since our D3-brane ends on the D5'-brane and is extended to $x^8=+\infty$, we should impose ${\rm Im}\,z >0$ on the moduli parameter $z$ so that the strings between the D5 and D3-branes can saturate the BPS bound.
\footnote{
If we set ${\rm Im}\,z<0$ with $B=0$, a string stretched between the D5 and D3-branes cannot be orthogonal to the D3-brane, and therefore breaks all the supersymmetry.
%If we set ${\rm Im}\,z<0$ with $B=0$, a string stretched between the D5 and D3-branes cannot saturate the BPS bound. 
%To saturate the BPS bound with charges fixed, it should be stretched between the D5 and D5'-branes with or without the winding number along $x^9$-direction. 
%Since such a string is bound to the D5-branes rather than the D3-brane, it is dual to the D2-D0 states in the original type IIA setup. They are not of interest in this paper. We will discuss the case of $B\neq 0$ in section \ref{sec:wall-crossing}.
} 
Such strings generally have the winding number $n$ along the compactified $S^1$ and the electric charge $\pm 1$ on the D5-brane. These two charges correspond to the D0 and D2-brane charges in the original type IIA setup. From Table \ref{table:D5-D3-F1}, we find that such open strings have $\nu=8$ and therefore the lowest energy states for the strings arise from the R-sector. This means that the BPS strings between the D5 and D3-branes are fermions in target space. In fact, the R-sector has two fermionic zero modes for oscillations in $x^0$ and $x^9$-directions (see Table \ref{table:D5-D3-F1}), which form the two-dimensional Clifford algebra
\begin{eqnarray}
 \left\{\psi_0^\mu,\psi_0^\nu\right\} &=& \eta^{\mu\nu},\label{eq:Clifford_alg}
\end{eqnarray}
for $\mu,\nu = 0,9$. Thus, before the GSO-projection, the lowest energy states behave as a two-dimensional Dirac fermion in target space, while the GSO-projection reduces a chiral half of them. The chirality operator is now defined by
\begin{eqnarray}
 \Gamma &=& \Gamma^0\Gamma^9 \;=\;2\psi_0^0\psi_0^9,
\end{eqnarray}
and we choose the GSO-projected lowest energy state $\left|k^0, n\right>$ so that $\Gamma\left|k^0,n\right> = -\left|k^0,n\right>$. Here, the charge $k^0$ denotes the energy of the BPS string, while $n$ is its winding number along $x^9$-direction. Note here that there are no other conserved momentum or winding number in $x^1,\cdots, x^8$ directions. For later use, we here define $k^9$ in terms of the winding number $n$ by
\begin{eqnarray}
 k^9 = \frac{1}{2\alpha'}\left(\pm l_0 + 2nR\right).\label{eq:k^9}
\end{eqnarray}
Here, the sign $\pm$ depends on the choice of the string orientation, and $\pi l_0$ is the distance between the D5 and D3-branes. Recall also that $R$ is the radius of $S^1$ on which $x^9$-direction is compactified (see Fig.~\ref{fig:setup}). 

In order to obtain physical ground states, we have to further impose $L_0$ and $G_0$-conditions. In our notation, $L_0$ and $G_0$ are given by\footnote{We here adopt the so-called old covariant quantization, and therefore $L_0$ and $G_0$ does not include contributions from ghosts in \eqref{eq:L_0-G_0}. We also mention that the previous definition \eqref{eq:k^9} of $k^9$ is just for simplicity of equation \eqref{eq:L_0-G_0}.}
\begin{eqnarray}
L_{0}
%%%^{M}
=\alpha'k_\mu k^\mu+\cdots, \qquad G_{0}
%%%^{M}
=\sqrt{2\alpha'}k_{\mu}\psi_{0}^{\mu}+\cdots,\label{eq:L_0-G_0}
\end{eqnarray}
where $\mu = 0,9$ and the dots involve the contributions from oscillations. Hence, the $L_0$-condition implies
%%%From (\ref{lcondition}), we can easily find the relation
%\[
%-(k^{0})^{2}+(k^{9})^{2}=0
%\]
\begin{equation}
%%% k^{9}=\pm |k^{0}|,
(k^0)^2 = (k^9)^2,\label{eq:L_0}
\end{equation}
while the $G_0$-condition gives a constraint
\begin{eqnarray}
0 &=& \left(k_0 \psi_0^0 + k_9 \psi_0^9\right)\left|k^0,n\right>
\;\,=\,\; \left(-k^0 + k^9\right)\psi_0^0\left|k^0,n\right>.\label{eq:G_0}
\end{eqnarray}
In the second equality of \eqref{eq:G_0}, we used the condition $\Gamma\left|k^0,n\right> = -\left|k^0,n\right>$ and the anti-commutation relation \eqref{eq:Clifford_alg}. Thus, the $L_0$ and $G_0$-conditions imply that $k^0 = k^9$.
%%% Further (\ref{gcondition}) leads to
%%% \begin{eqnarray}
%%%( k_0 \Gamma^0 + k_9 \Gamma^9)|-,k\rangle_{R}&=&0,\nonumber \\
%%%\therefore \ \ k^0 &=&k^9 \label{momentum1}
%%% \end{eqnarray}
%%%
 %%%%%
%%% \begin{comment}
%%%\[
%%%\sqrt{\alpha'}(k_{0}\Gamma^{0}+k_{9}\Gamma^{9})|-,k\rangle_{R}=0,
%%%\]
%%%\[
%%%\left( \begin{array}{cc}
%%%0&-k_{0}+k_{9}\\
%%%k_{0}+k_{9}&0\\
%%%\end{array}\right)
%%%\left( \begin{array}{c}
%%%1\\
%%%0\\
%%%\end{array}\right)=0,
%%%\]
%%%\begin{equation}
%%%\therefore k^{0}=k^{9}
%%%\label{momentum1_1}
%%%\end{equation}
%%%\end{comment}
%%%%%%%%%
%%%To understand the meaning of this constraint, we recall that the 9-direction has the DD-type boundary condition,
%%%\begin{equation}
%%%X^{9}|_{\sigma=0}=0,\ \ \ X^{9}|_{\sigma=\pi}=\pi l=\pi(l_{0}+2n R)
%%%\end{equation}
%%%where $n=0,1,2,\cdots$ represents the winding number (see Fig\ref{fig:35}).
%%%By substituting $\bar{x}_{1}^{9}=0, \bar{x}_{2}^{9}=\pi l$ into (\ref{dd}), we obtain the expression of $k^9$ in terms of the parameters of our system
%%%\begin{equation}
%%%k^{9}=\frac{l}{2\alpha'}=\frac{1}{2\alpha'}(l_{0}+2nR)\ \ \ (n=0,1,2\cdots).
%%%\end{equation}
%%%Combining the above relation with (\ref{momentum1}), we have
%%%\begin{equation}
%%%k^{0}=\frac{1}{2\alpha'}(l_{0}+2nR)\ \ \ (n=0,1,2\cdots).
%%%\end{equation}
%%%This means that the winding number $n$ is determined uniquely for a given energy $k_{0}$.
This means that the energy of a BPS string stretched between the D5 and D3-branes is determined by its winding number $n$. Hence, we can conclude that {\em there is only one fermionic quantum state for each winding number $n$ of the BPS string stretched between the D5 and D3-branes.}
%%%Taking into account that this state corresponds to a space-time fermion, we can conclude that \textit{we have one fermionic BPS state from the open string between D3 and D5 branes.}

% subsection 1.3

Note here that equation \eqref{eq:k^9} and the condition $k^9 = k^0 \geq 0$ imply that the winding number should satisfy $n\geq 0$ or $n>0$ according to the orientation of the string. Changing the string orientation corresponds to the charge conjugation for the original D2-branes in the type IIA side. Therefore the winding number $n$ is positive or zero if the original D2-brane charge $Q_2=1$, while it should be positive if $Q_2=-1$. This fact is important when we evaluate the BPS partition function. Hereafter, we call strings for $Q_2=1$ ``D3-D5 strings'' and strings for $Q_2=-1$ ``D5-D3 strings'' (see Fig.~\ref{fig:three}).
%%%\subsection{D5-D3 string}\label{d5-d3}
%%%In the case of the D5-D3 string, the procedure is almost same, but the boundary condition is different from that of the D3-D5 string:
%%%\begin{equation}
%%%X^{9}|_{\sigma=0}=\pi l+2nR,\ \ \ X^{9}|_{\sigma=\pi}=0
%%%\end{equation}
%%%where $n=1,2,\cdots$ denotes winding number (see Fig\ref{fig:53}).
%%%Note that $n=0$ is excluded in this case because it is included in the case of the D3-D5 string.
%%%Corresponding to this modification, we obtain
%%%\begin{equation}
%%%k^{9}=-\frac{l}{2\alpha'}=-\frac{1}{2\alpha'}(l_{0}+2nR)\ \ \ (n=1,2,\cdots).
%%%\end{equation}
%%%In short, we see that \textit{we have one fermionic BPS state from the open string  between D5-D3 branes.}

%subsection1.4

\subsection{Open strings on the D3-brane}
\label{subsec:D3-D3}

We here investigate strings on the D3-brane, namely ``D3-D3 strings.''
With the vanishing winding number, such strings just describe the quantum fluctuations of the D3-brane. But now, since $x^9$-direction is compactified on $S^1$, there are BPS strings on the D3-brane with the non-vanishing winding number $n$ along the $S^1$. The winding number $n$ corresponds to the D0-brane charge $Q_0$ in the type IIA side. For the non-vanishing $Q_0 = n$, this system again becomes a D5-D3-F1 system, but now it does not have any D2-brane charge. This is because both ends of such strings are on the D3-brane. Hence, these BPS bound states are dual to the D4-D0 states on the resolved conifold in the type IIA side.

Let us now count the number of such BPS strings. For each value $n$ of the winding number there is a supersymmetric string ground state\footnote{Counting the states of this contribution includes subtlety; this D3-brane includes non-compact direction and the number of states depends on the boundary condition. However we here naively assume that the ``Euler number'' of this non-compact space is $1$.  This is just the same subtlety appeared in the D4-D0 bound states in IIA side.}, which follows from \eqref{eq:NS_zero} and \eqref{eq:Ramond_zero}. Such a single-string configuration contributes $u^n$ to the BPS partition function. However, there are more string configurations that give $u^n$ to the partition function. For example, let us consider two strings on the D3-brane with the winding numbers $n-1$ and $1$, respectively. Such a two-string configuration also contributes $u^n$ to the BPS partition function.
In general, for a given value $n$, there are $p(n)$ configurations of strings on the D3-brane that contribute $u^n$ to the BPS partition function, where $p(n)$ denotes the partition number of $n$. Then, we can write a generating function of such BPS bound states as\footnote{If there is no D3-brane here, then such multi-string configurations have no binding energy and the constituents are not bound to each other. In that case, we should not regard them as bound states. But we now have a D3-brane, and the strings are bound to the brane. This implies that the BPS multi-string configurations on the D3-brane can be seen as BPS bound states, and we should count all of them to evaluate the generating function.}
\begin{eqnarray}
\sum_{n=0}^\infty p(n)u^n 
&=&
\prod_{n=1}^\infty\left(
\frac{1}{1-u^{n}}
\right).\label{eq:D3-D3}
\end{eqnarray}
This is precisely equal to the BPS partition function of D0-branes bound to a single D4-brane on $\mathbb{C}^2$.

%subsection 1.5

\subsection{BPS partition function}

Having discussed the number of BPS
%%%D3-D5-D5'
D5-D3-F1 bound states, we can now calculate the BPS partition function.
%%%of it.
 The result we will obtain here is consistent with \cite{Nishinaka:2010qk}, which supports the validity of the TS-duality of the conifold.

% section1.5.1
%%%
%%%\subsubsection{BPS Partition Function}
%%%We assign the following charges
%%%\begin{equation}
%%%\begin{tabular}{|c||c|cc|} \hline 
%%%charge & D4-D2-D0 system& D3-D5-D5' system& \\ \hline \hline
%%%$Q_{0}$&$\textrm{D}0\textrm{charge}$ &$\textrm{winding number of F1}$&$n$\\ \hline
%%%$Q_{2}$&$\textrm{D}2\textrm{ charge}$ &$\textrm{D}3\textrm{-brane} \  U(1)\textrm{gauge charge}$& $\pm 1,0$ \\ \hline
%%%$P_{2}$&$\textrm{D}4\textrm{charge}$&$1$& \\ \hline
%%%$P_{0}$&$\textrm{D}6\textrm{charge}$&$0$& \\ \hline
%%%\end{tabular}
%%%\end{equation}
To evaluate the BPS partition function, we use the following facts:
\begin{enumerate}
\item BPS strings between the D5 and D3-branes with the winding number $n$ carry the D0-brane charge $Q_0 = n$. The possible values of $n$ depends on their D2-brane charge $Q_2$, that is, $n\geq 0$ for $Q_2=1$ while $n>0$ for $Q_2 = -1$. We call strings for $Q_2 = 1$ ``D3-D5 strings'' and those for $Q_2=-1$ ``D5-D3 strings.''
%%%$Q_{2}=+1, 0, -1$ respectively corresponds to the case of the D3-D5, D5-D3, D3-D3 string(see Fig. \ref{fig:three}).
\item Strings on the D3-brane carry the vanishing D2-brane charge. Their contributions to the partition function has been evaluated as \eqref{eq:D3-D3}.
\item The number of fermionic BPS states with charge $\gamma$ contributes to the Witten index $\Omega(\gamma)$ with a minus sign.
%%%The Witten index $\Omega(\gamma)$ for the fermions gives a minus sign in the BPS partition function. 
\end{enumerate}
Combining these facts,
%%%with the previous results in \ref{subsec:D5-D3} and \ref{subsec:D3-D3},
%%%Section \ref{d3-d5}, \ref{d5-d3}, \ref{d3-d3},
we can evaluate the BPS partition function \eqref{eq:partition_function} in our type IIB setup.
%%%determine contributions from each type of the open strings to the partition function.
Let us first evaluate the contribution from the D3-D5 strings. Such strings have the winding number $n=0,1,2,\cdots$, each of which contributes $-u^nv$ to the partition function. Recall here that BPS D3-D5 strings are spacetime fermions. In general, we can consider multi-string configurations of such fermionic D3-D5 strings, and therefore the total contributions of the D3-D5 strings to the partition function can be written as
\begin{eqnarray}
%%%(1-v)(1-uv)(1-u^{2}v)\cdots=
\prod_{n=0}^{\infty}(1-u^{n}v).
\end{eqnarray}
Similarly, we can evaluate the contributions of the D5-D3 strings. The only one difference is that the winding number $n$ now runs over $n=1,2,3,\cdots$. Thus, the D5-D3 strings contribute
\begin{eqnarray}
%%%(1-uv^{-1})(1-u^{2}v^{-1})\cdots =
\prod_{n=1}^{\infty}(1-u^{n}v^{-1})
\end{eqnarray}
to the partition function.
Note that the D5-D3 strings have the different orientation from the D3-D5 strings, and therefore an opposite D2-brane charge $Q_2=-1$.
%%%\begin{eqnarray}
%%%(1+u+u^{2}+\cdots)(1+u^{2}+u^{4}+\cdots)=
%%%\prod_{n=1}^{\infty}\frac{1}{1-u^{n}}.
%%%\end{eqnarray}

Then, taking these together with the contributions \eqref{eq:D3-D3} from the D3-D3 strings, we find that the full BPS partition function is evaluated as
\begin{equation}
\mathcal{Z}_{\rm BPS}(u,v) = \prod_{k=1}^{\infty}
\left(\frac{1}{1-u^k}\right)\prod_{m=0}^\infty(1-u^{m}v)\prod_{n=1}^\infty(1-u^{n}v^{-1}).
\label{zbpseq}
\end{equation}
This result agrees with the D4-D2-D0 partition function \eqref{bps1} for $n_0=\infty$. The limit $n_0=\infty$ means that all the walls of marginal stability are crossed when we move the moduli $z$ from ${\rm Im}\,z = 0$ to ${\rm Im}\,z>0$. The reason for this is that we now do not take into account the B-field $B$ on the original D4-brane in the type IIA side. By taking it into account, in the next subsection, we will see more concrete correspondence between the type IIA and IIB sides.
%%%, which supports the validity of the TS-duality transformation on the conifold.
%%%Therefore, it gives us one of the evidence for the TS-duality.

%subsection1.6

\begin{comment}
\subsection{Insertion of electromagnetic field}
In the subsection 1.5, we showed that BPS partition function could be derived from dualized D3-D5-D5' system by using string theory. This result is interesting itself, however moreover we can obtain more fascinating result. That is the visualization of "wall crossing phenomena".
 This analysis will give us stringy pictorial interpretation and  make it clear to understand what happens when we encounter the wall crossing phenomena in string theory. To understand it, we have to insert electric field along 8 direction into this system.
So we start by discussing the effect of inserting electromagnetic field.
\end{comment}

%subsubsection1.6.1

\section{Wall-crossing phenomena in type IIB side}
\label{sec:wall-crossing}
%\subsubsection{Wall Crossing Phenomena}

%The explicit expressions of wall crossing phenomena of the D4-D2-D0 system on the conifold\cite{Nishinaka:2010qk},
%\begin{equation}
%Z_{BPS}=(1-v)\prod_{n=1}^{\infty}
%\frac{1}{1-u^{n}}\prod_{m=1}^{\infty}(1-u^{m}v)
%\label{bpseq2}
%\end{equation}
%\begin{equation}
%Z_{BPS}=\prod_{n=1}^{\infty}
%\frac{1}{1-u^{n}}.
%\label{bpseq3}
%\end{equation}
We now examine how the wall-crossing phenomena can be understood in the type IIB side.
%We have shown the dualized D3-D5-D5' system correctly counts the BPS partition function of the original D4-D2-D0 system.
For this purpose, let us turn on the B-field along the non-compact cycles in the original type IIA side. As seen in subsection \ref{subsec:moduli}, the B-field on the divisor wrapped by the D4-brane can be mapped to the electric field $F_{08}$ on the dual D3-brane. Such an electric field modifies the BPS condition of our D5-D3-F1 system, which leads to different ``slope'' of the fundamental strings stretched between the D5 and D3-branes.
We will see that
such a modified IIB picture provides
%it also gives an interesting
a pictorial interpretation of the wall crossing phenomena of our BPS states.
%To see this, we first discuss about the open string states under the existence of the electric field along $F_{08}$, which arises from the phase of the K\"ahler moduli associated with the non-compact cycles of the conifold. We then study, for a fixed $F_{08}$, how the BPS partition function varies as we change the moduli $t$, which is the distance between D5 and D5' branes.

%subsubsection1.6.1

\subsection{Effect of electric field}\label{electricsec}

Let us consider the BPS configuration of strings between the D5 and D3-branes under the non-vanishing electric field $F_{08}$ on the D3-brane. Recall that the world-sheet action of a string coupled to the electric field at its boundary is written as\footnote{We here write only the bosonic part for simplicity.}
%%%Under the existence of an electromagnetic field $A_{m}$ on the D$p$-brane, the open string action is given by
\begin{equation}
S= \frac{1}{4\pi\alpha'}\int d\tau d\sigma(\partial_{a}X^{\mu}\partial_{a}X_{\mu})
+\int d\tau A_{m}(X)\frac{dX^{m}}{d\tau}\biggl|_{\sigma=0}\label{action2},
\end{equation}
where $a=\tau,\sigma$ and $\mu=0,\cdots,9$. We set $\sigma=0$ to be the boundary of the string ending on the D3-brane, and therefore $m$ takes values in $m=0,6,7,8$, i.e.\ the directions in which our D3-brane is extended. 
%%%Suppose that $F_{mn}$ is  constant, then we can choose the following gauge
For a constant background field $F_{08}$, the gauge potential $A_m$ can be written in a suitable gauge as
\begin{equation}
A_{m}(X)=\frac{1}{2}F_{mn}X^{n}.
\end{equation}
%%%In this gauge, the action (\ref{action2}) becomes 
%%%\begin{equation}
%%%S= \frac{1}{4\pi\alpha'}\int d\tau d\sigma(\partial_{a}X^{\mu}\partial_{a}X_{\mu})
%%%+\int d\tau \frac{1}{2} F_{nm}X^n\frac{dX^{m}}{d\tau}\biggl|_{\sigma=0}^{\sigma=\pi}.
%%%\end{equation}
The variation of the world-sheet action with respect to $X^{m}$ leads to the boundary condition,\footnote{We impose the Dirichlet boundary condition on $X^1,\cdots,X^5$ and $X^9$ at $\sigma=0$. For the other boundary $\sigma=\pi$, we impose the Neumann boundary condition on $X^0,\cdots,X^5$ and the Dirichlet boundary condition on $X^6,\cdots,X^9$, corresponding to the existence of the D5-brane.}
\begin{equation}
\left.\left(\partial_{\sigma}X_{m}-2\pi\alpha' F_{mn}\partial_{\tau}X^{n}\right)\right|_{\sigma = 0} = 0,
\end{equation}
%%%for $\sigma=0, \pi$.
which particularly implies 
\begin{eqnarray}
 \left.\partial_\sigma X^8\right|_{\sigma=0} = -\left.2\pi\alpha'F_{08}\,\partial_\tau X^0\right|_{\sigma=0}.
\label{eq:boundary_condition}
\end{eqnarray}

From the condition \eqref{eq:boundary_condition}, we find that our BPS strings end on the D3-brane with a ``slope'' depending on the electric field $F_{08}$. To see this explicitly, we estimate $\delta X^8/\delta X^9$ along the string world-sheet. Note that our D5-D3 strings cannot have non-vanishing momentum in the $x^9$-direction. This implies that for BPS D5-D3 strings $X^9(\tau,\sigma)$ has no $\tau$-dependence.\footnote{For BPS strings, as mentioned before, all the string excitations vanish so that the strings have the lowest energy. For such strings, $X^\mu$ depends on $\tau$ if and only if it has non-vanishing momentum $p^\mu$.} Therefore, we can identify $\delta X^8/\delta X^9$ with
$
\partial_\sigma X^8 / \partial_\sigma X^9
$
for BPS D5-D3 strings.
In particular, taking the Virasoro constraint $T_{ab}=0$ into account, this slope can be evaluated at the boundary $\sigma=0$ as
%\begin{eqnarray}
%  \left.\frac{\delta X^8}{\delta X^9}\right|_{\sigma=0} = -\left.2\pi\alpha'F_{08}\,\frac{\partial_\tau X^0}{\partial_\sigma X^9}\right|_{\sigma=0}.
%\label{eq:slope-condition}
%\end{eqnarray}
%Note that $\partial_\tau X^0$ corresponds to the energy of the BPS strings and does not vanish.\footnote{The denominator $\partial_\sigma X^9$ expresses the winding number of the BPS strings. Since our D5 and D3-branes are separated with a distance $\pi l_0$ (ee Fig. \ref{fig:setup}), this quantity cannot vanish. Therefore, the right-hand side of \eqref{eq:slope-condition} is always well-defined.} Thus, it follows that $\left.\delta X^8/\delta X^9\right|_{\sigma=0}$ vanishes if and only if $F_{08}=0$, and otherwise $\left.\delta X^8/\delta X^9\right|_{\sigma=0}$ takes a non-zero value. In fact, from the Virasoro constraint $T_{ab}=0$ at the end point one finds the value of this slope is given by
\begin{align}
 \left.\frac{\delta X^8}{\delta X^9}\right|_{\sigma=0} =\frac{-2\pi\alpha'F_{08}}{\sqrt{1-(2\pi \alpha' F_{08})^2}}=:-\tan\theta.
\label{eq:slope-condition}
\end{align}
This quantity vanishes when $F_{08}=0$, while does not vanish when $F_{08}\ne 0$\footnote{The inequality $|2\pi \alpha' F_{08}|\le 1$ must be satisfied}.

\begin{figure}
\begin{center}
\includegraphics[width=12cm]{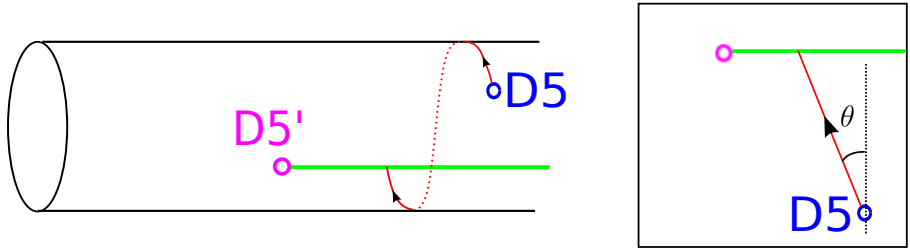}
\caption{The sketch of a BPS D5-D3 string, which is tilted by the electric field $F_{08}$ in the $x^8$-$x^9$ plane. The green line stands for the D3-brane. The ``slope'' of the D5-D3 string $\theta$ depends on the electric field $F_{08}$ on the D3-brane. If $F_{08}=0$, then $\theta$ vanishes.}
\label{fig:add_E}
\end{center}
\end{figure}

The quantity $\delta X^8/\delta X^9$ represents the ``slope'' of the BPS strings in $x^8$-$x^9$ plane. If it vanishes, the strings wind along the $x^9$-direction and localized in $x^8$-direction. This is the case for $F_{08}=0$, which has already been analyzed in the previous section. On the other hand, if $F_{08}\neq0$ then the slope $\delta X^8/\delta X^9$ takes a non-zero value, at least at the string boundary on the D3-brane. Since our strings saturate the BPS bound and have no excitations, they should have the constant slope along their world-sheets. This implies that our BPS strings are stretched between the D5 and D3-branes {\em with a constant slope depending on $F_{08}$.} The typical example for some non-vanishing $F_{08}$ is depicted in Fig.~\ref{fig:add_E}. In the rest of this paper, we set $F_{08}\neq 0$ so that $\delta X^8/\delta X^9 < 0$ is satisfied.\footnote{This corresponds to the condition $\pi/4<\varphi<\pi/2$ imposed in \cite{Nishinaka:2010qk}.}

We here briefly mention the D2-brane charge $Q_2$ in the type IIA side. As seen at the end of subsection \ref{subsec:D5-D3}, the winding number along $x^9$-direction is non-negative for both the D5-D3 strings ($Q_2<0$) and the D3-D5 strings ($Q_2>0$). This and the fact that $\delta X^8/\delta X^9<0$ imply that the string endpoint of $\sigma=\pi$ is always located on the left of the other endpoint $\sigma=0$, for both the D5-D3 and D3-D5 strings. Therefore, it follows that {\em D5-D3 strings are extended to the left of the D5-brane while the D3-D5 strings are extended to the right of it.} This fact will be very important when we examine how the results in the previous section are modified under the non-vanishing $F_{08}$.

%subsection1.7
\subsection{Wall crossing phenomena}
%Now we are now ready to see the wall crossing phenomena from the view of the open string spectrum. 
%As discussed previously, the open string becomes tilted in the 8-9 plane by the electric background $F_{08}$. 
%figure3 from Okada-san
%For a nonzero $F_{08}$, we have the open strings tilted with the angle $\theta$ in the 8-9 plane. 
We are now ready to see how the wall-crossing phenomena can be understood in the presence of the non-vanishing electric field $F_{08}$ on the D3-brane. We count the number of BPS strings on the D5-D3 system, as in section \ref{sec:open-string}. The only difference from section \ref{sec:open-string} is the ``slope'' of the strings induced by the electric field $F_{08}\neq 0$.
%%%, for a nonzero $F_{08}$, how the wall crossing in the moduli parameter $z$, the distance between D5-brane and D5'-brane, can be understood from the view of the open string spectrum.

\begin{figure}
\begin{center}
\includegraphics[width=6cm]{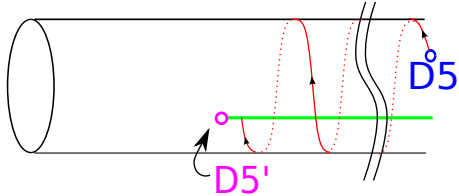}
\caption{The sketch of the D5-D3 string in  case of ${\rm Im}\,z=\infty$, where D3-D5 and D5-D3 strings with arbitrary winding numbers exist in the BPS spectrum.}
\label{fig:positive_infinite}
\end{center}
\end{figure}

Let us first consider the limiting case of ${\rm Im}\,z=+\infty$. Recall that ${\rm Im}\,z$ represents the relative position of the D5-brane in $x^9$-direction. So, the limit ${\rm Im}\,z=+\infty$ corresponds to moving the D5-brane far right in Fig.~\ref{fig:positive_infinite}.
Then, all the BPS D5-D3 and D3-D5 strings considered in the previous section still exist, and the corresponding BPS partition function is the same as \eqref{zbpseq}, that is,
\begin{eqnarray}
 \mathcal{Z}_{+\infty}(u,v) &=& \prod_{k=1}^{\infty}\left(\frac{1}{1-u^k}\right)\prod_{m=0}^\infty(1-u^{m}v)\prod_{n=1}^\infty(1-u^{n}v^{-1}).
\end{eqnarray}

\begin{figure}
\begin{center}
\includegraphics[width=6cm]{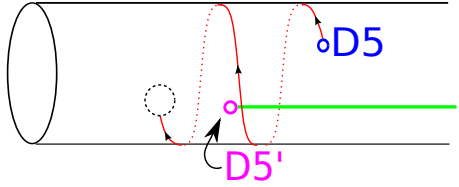}
\caption{For ${\rm Im}\,z>0$, some of the D5-D3 strings with large winding numbers does not exist in the BPS spectrum, while all the D3-D5 strings are stable. The maximum winding number $n_0$ of the D5-D3 strings depends on the moduli ${\rm Im}\,z$.}
\label{fig:positive}
\end{center}
\end{figure}

Next, let us move the moduli parameter ${\rm Im}\,z$ away from ${\rm
Im}\,z = +\infty$ while fixing ${\rm Re}\,z$ as $0< {\rm Re}\, z< 1$.\footnote{For example ${\rm Re}\,z=1/2 $ in \cite{Nishinaka:2010qk}.}
%from $+\infty$ to $-\infty$.
% For the case of $t=+\infty$, the BPS partition function is given by (\ref{zbpseq})
%\[
%Z_{BPS}=(1-v)\prod_{n=1}^{\infty}
%\frac{1}{1-u^{n}}\prod_{m=1}^{\infty}(1-u^{m}v)(1-u^{m}v^{-1}).
%\]
%This case includes all the possible BPS states (see Fig.\ref{fig:positive_infinite}).
For some finite value of ${\rm Im}\,z>0$, all the D3-D5 strings still remain stable, but some D5-D3 strings disappear from the spectrum.
%%%we lose contributions to the partition function from the D5-D3 strings whose winding numbers are less than some critical value, which is determined by the ${\rm Im}\,z$.
To see this, one considers the BPS D5-D3 string which starts from the D5-brane and extends to the left in Fig.~\ref{fig:positive}. Such a D5-D3 string is stretched between the D5 and D3-branes with the fixed slope determined by $F_{08}\neq 0$. This implies that the string endpoint on the D3-brane moves to the left in Fig.~\ref{fig:positive} as its winding number increases.
However, since our D3-brane ends at the D5'-brane, there is a maximum value $n_0$ of the possible winding number. If a D5-D3 string has the winding number larger than $n_0$, the ``left'' endpoint of the string can not reach the D3-brane while keeping the BPS condition. Therefore, such a D5-D3 string does not exist in the BPS spectrum (see Fig.~\ref{fig:positive}). Thus, we find that the stable BPS D5-D3 string exists if and only if its winding number $n$ is less than or equal to $n_0$.

\begin{figure}
 \begin{center}
\includegraphics[width=6cm]{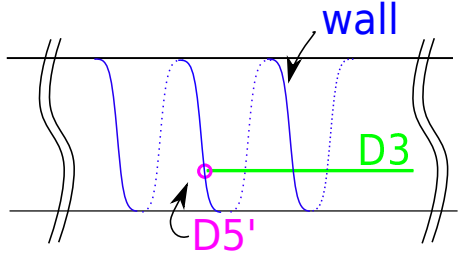}
\end{center}
\caption{The moduli space and the walls of
 marginal stability.}
\label{fig:wallb}
\end{figure}

Note here that the maximum winding number $n_0$ of the D5-D3 strings
depends on the moduli ${\rm Im}\,z$. If we keep decreasing ${\rm
Im}\,z$, the maximum winding number $n_0$ decreases by one at ${\rm
Im}\,z=(n- {\rm Re}\, z)\tan\theta,\ n\in \mathbb{Z}$ with $\theta$ in
eq.\eqref{eq:slope-condition}, which means that one D5-D3 string becomes
unstable and disappears from the spectrum at each such value of ${\rm
Im}\,z$. Such special values of ${\rm Im}\,z$ are on the walls of
marginal stability (see Fig.~\ref{fig:wallb}). At each such value of ${\rm Im}\,z$, the BPS D5-D3 string with winding number $n_0$ can marginally decay into a D5-D5' string.\footnote{ This corresponds, in the original type IIA side, to the separation of a D2-D0 fragment from a D4-D2-D0 bound state.} This is nothing but the wall-crossing phenomenon!
Let us call the chamber expressed by
\begin{align}
 (n_0- {\rm Re}\, z)\tan\theta<{\rm Im}\,z<(n_0+1- {\rm Re}\, z)\tan\theta 
\end{align}
the ``$n_0$-th chamber.''

\begin{figure}
\begin{center}
\subfigure[No D5-D3 string exists.]{\includegraphics*[width=.35\linewidth]{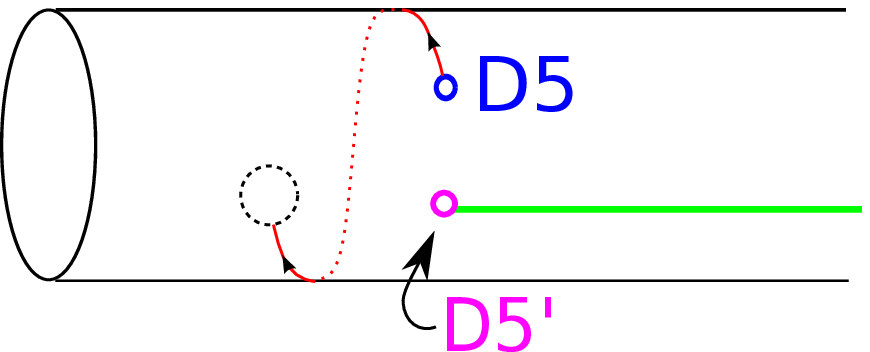}}\qquad\qquad
\subfigure[The D3-D5 string without winding can exist.]{\includegraphics*[width=.35\linewidth]{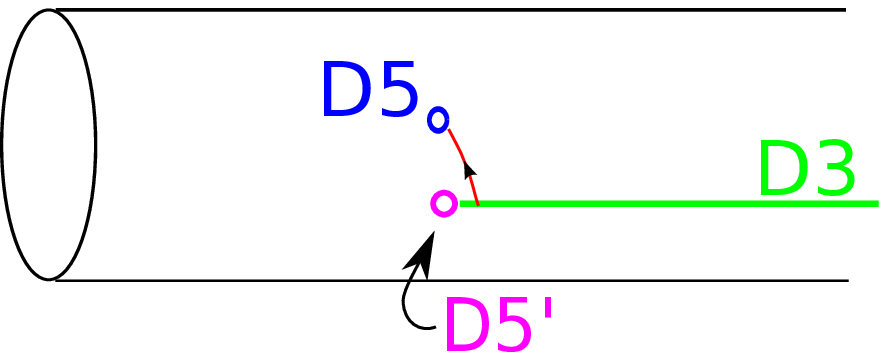}}
\end{center}
\caption{At ${\rm Im}\,z=0$, we have no D5-D3 strings. So all we have are D3-D5 strings and D3-D3 strings.}
\label{fig:zero}
\end{figure}

When ${\rm Im}\,z>0$, only D5-D3 strings have such wall-crossing phenomena and all the D3-D5 strings are always stable. Therefore, the BPS partition function is written in the $n_0$-th chamber ($n_0>0$) as
\begin{eqnarray}
 \mathcal{Z}_{n_0}(u,v) = \prod_{k=1}^\infty\left(\frac{1}{1-u^k}\right)\prod_{m=0}^\infty(1-u^mv)\prod_{n=1}^{n_0}(1-u^nv^{-1}).
\end{eqnarray}
In the $0$-th chamber, all the D5-D3 strings become unstable (see the left picture of Fig.~\ref{fig:zero}), and the corresponding BPS partition function becomes 
\begin{eqnarray}
\mathcal{Z}_{0}=\prod_{k=1}^{\infty}
\left(\frac{1}{1-u^k}\right)\prod_{m=0}^\infty(1-u^{m}v).
\end{eqnarray}
These results perfectly agree with equation \eqref{bps1} which was obtained by the analysis in the original type IIA side.\footnote{$\theta$ in this paper is related to $\varphi$ in \cite{Nishinaka:2010qk} as $\pi-\theta=2\varphi$. }

\begin{figure}
\begin{center}
%\subfigure[For some negative $t<0$.]{\includegraphics*[width=.3\linewidth]{decreasing_negative.eps}
%\label{fig:negative}}
\subfigure
%[For the smaller $t<0$.]
{\includegraphics*[width=.4\linewidth]{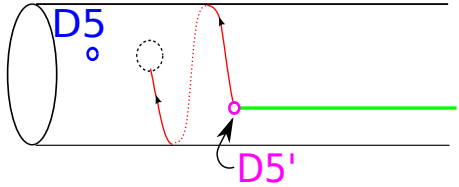}}
\end{center}
\caption{For ${\rm Im}\,z<0$, there is a minimum value of the winding numbers of D3-D5 strings. The minimum winding number $m_0$ depends on the moduli ${\rm Im}\,z$.}
\label{fig:negative}
\end{figure}

\begin{figure}
\begin{center}
\includegraphics[width=6cm]{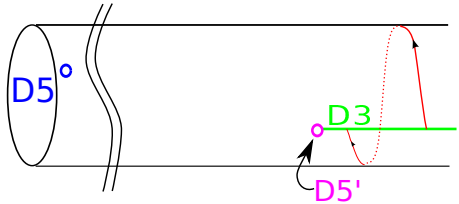}
\caption{For ${\rm Im}\,z=-\infty$, only the D3-D3 strings exist in the spectrum.}
\label{fig:negative_infinite}
\end{center}
\end{figure}

%figure5 from Okada-san
Let us next consider the case of negative ${\rm Im}\,z$. When we keep decreasing ${\rm Im}\,z$ in the region of ${\rm Im}\,z<0$, some of the BPS D3-D5 strings might, in turn, be unstable and disappear from the spectrum at some values of ${\rm Im}\,z$. This is because,  when ${\rm Im}\,z$ is negative,
the D5-brane is located on the left side of the D5'-brane in Fig.~\ref{fig:negative}, which implies that there is a minimum value $m_0$ of the winding number of the BPS D3-D5 strings. Actually $m_0=|n_0|$ in the $n_0$-th chamber ($n_0<0$).\footnote{Note that all the D5-D3 have already disappeared from the spectrum.}
%%%as we decrease ${\rm Im}\,z<0$, the D5-brane moves away from the D5'-brane, and some of strings staring the D3-brane can not reach the D5-brane (see Fig.\ref{fig:negative}).
Thus, the corresponding BPS partition function in the $n_0$-th chamber ($n_0<0$) is now given by
\begin{eqnarray}
 \mathcal{Z}_{n_0}(u,v) = \prod_{k=1}^\infty\left(\frac{1}{1-u^k}\right)\prod_{m=|n_0|}^\infty(1-u^mv).
\end{eqnarray}
%The minimum winding number $m_0$ for the D3-D5 strings depends on the moduli ${\rm Im}\,z<0$. 
In particular, in the limit of ${\rm Im}\,z=-\infty$, the partition function has no $v$-dependence:
\begin{eqnarray}
\mathcal{Z}_{-\infty}(u,v) = \prod_{k=1}^{\infty}
\left(\frac{1}{1-u^{k}}\right).
\end{eqnarray}
This is because there is no BPS open string stretched between the D5 and D3-branes while only the D3-D3 strings remain stable (see Fig.~\ref{fig:negative_infinite}).
These results again completely agree with (\ref{bps2}) which was obtained from the type IIA analysis.

\begin{figure}[h!]
 \begin{center}
\includegraphics[width=7cm]{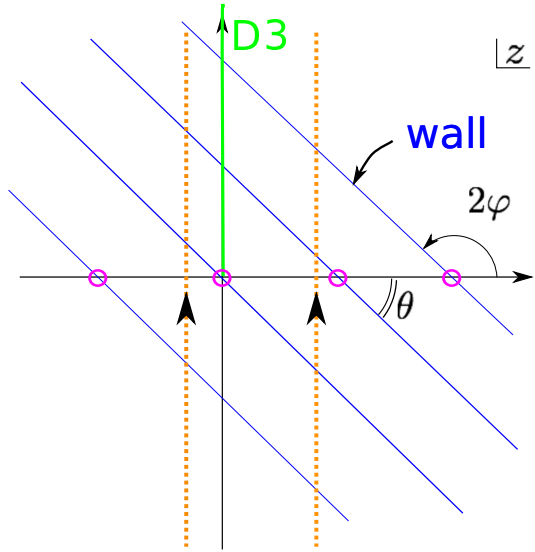}
\end{center}
\caption{The universal covering of the moduli space and the walls of marginal stability obtained from Fig.~\ref{fig:wallb}. Two dotted lines are identified. 
This picture exactly coincides with \cite{Nishinaka:2010qk}.}
\label{fig:walla}
 \end{figure}

All the above results perfectly agree with those obtained by the
analysis in the original type IIA side, that is, the wall-crossing
phenomena of D4-D2-D0 states on the conifold.
 Namely we obtain the same structures of the walls of marginal stability
 (see Fig.~\ref{fig:walla})
 and the same  partition functions in each chamber in both sides.
 This agreement strongly
supports the validity of still mysterious S-duality of string theory.
%%%%%%%%%%%%%%%%%%%%%%%%%%%%%%%%%%%
\subsection*{Acknowledgments}

We would like to thank Kazutoshi Ohta, Yuji Okawa, Takuya Okuda and Hiroaki Tanida for discussions and comments. 
S.Y. was supported in part by KAKENHI 22740165.
%%%%%%%%%%%%%%%%%%%%%%%%%%%%%%%%%%5
\providecommand{\href}[2]{#2}\begingroup\raggedright\endgroup


\begin{thebibliography}{10}

\bibitem{Denef:2000nb}
F.~Denef, ``{Supergravity flows and D-brane stability},'' {\em JHEP} {\bfseries
  08} (2000) 050,
\href{http://arxiv.org/abs/hep-th/0005049}{{\ttfamily arXiv:hep-th/0005049}}.
%%CITATION = HEP-TH/0005049;%%.

\bibitem{Denef:2007vg}
F.~Denef and G.~W. Moore, ``{Split states, entropy enigmas, holes and halos},''
\href{http://arxiv.org/abs/hep-th/0702146}{{\ttfamily arXiv:hep-th/0702146}}.
%%CITATION = HEP-TH/0702146;%%.

\bibitem{deBoer:2008fk}
J.~de~Boer, F.~Denef, S.~El-Showk, I.~Messamah, and D.~Van~den Bleeken,
  ``{Black hole bound states in AdS\verb|_|3 x S\verb|^|2},''
  \href{http://dx.doi.org/10.1088/1126-6708/2008/11/050}{{\em JHEP} {\bfseries
  11} (2008) 050},
\href{http://arxiv.org/abs/0802.2257}{{\ttfamily arXiv:0802.2257 [hep-th]}}.
%%CITATION = 0802.2257;%%.

\bibitem{deBoer:2008zn}
J.~de~Boer, S.~El-Showk, I.~Messamah, and D.~Van~den Bleeken, ``{Quantizing N=2
  Multicenter Solutions},''
  \href{http://dx.doi.org/10.1088/1126-6708/2009/05/002}{{\em JHEP} {\bfseries
  05} (2009) 002},
\href{http://arxiv.org/abs/0807.4556}{{\ttfamily arXiv:0807.4556 [hep-th]}}.
%%CITATION = 0807.4556;%%.

\bibitem{Cardoso:2008ej}
G.~L. Cardoso, J.~R. David, B.~de~Wit, and S.~Mahapatra, ``{The mixed black
  hole partition function for the STU model},''
  \href{http://dx.doi.org/10.1088/1126-6708/2008/12/086}{{\em JHEP} {\bfseries
  12} (2008) 086},
\href{http://arxiv.org/abs/0810.1233}{{\ttfamily arXiv:0810.1233 [hep-th]}}.
%%CITATION = 0810.1233;%%.

\bibitem{VanHerck:2009ww}
W.~Van~Herck and T.~Wyder, ``{Black Hole Meiosis},''
  \href{http://dx.doi.org/10.1007/JHEP04(2010)047}{{\em JHEP} {\bfseries 04}
  (2010) 047},
\href{http://arxiv.org/abs/0909.0508}{{\ttfamily arXiv:0909.0508 [hep-th]}}.
%%CITATION = 0909.0508;%%.

\bibitem{Andriyash:2010qv}
E.~Andriyash, F.~Denef, D.~L. Jafferis, and G.~W. Moore, ``{Wall-crossing from
  supersymmetric galaxies},''
\href{http://arxiv.org/abs/1008.0030}{{\ttfamily arXiv:1008.0030 [hep-th]}}.
%%CITATION = 1008.0030;%%.

\bibitem{Andriyash:2010yf}
E.~Andriyash, F.~Denef, D.~L. Jafferis, and G.~W. Moore, ``{Bound state
  transformation walls},''
\href{http://arxiv.org/abs/1008.3555}{{\ttfamily arXiv:1008.3555 [hep-th]}}.
%%CITATION = 1008.3555;%%.

\bibitem{Manschot:2010qz}
J.~Manschot, B.~Pioline, and A.~Sen, ``{Wall-Crossing from Boltzmann Black Hole
  Halos},''
\href{http://arxiv.org/abs/1011.1258}{{\ttfamily arXiv:1011.1258 [hep-th]}}.
%%CITATION = 1011.1258;%%.

\bibitem{Manschot:2011xc}
J.~Manschot, B.~Pioline, and A.~Sen, ``{A fixed point formula for the index of
  multi-centered N=2 black holes},''
  \href{http://dx.doi.org/10.1007/JHEP05(2011)057}{{\em JHEP} {\bfseries 05}
  (2011) 057},
\href{http://arxiv.org/abs/1103.1887}{{\ttfamily arXiv:1103.1887 [hep-th]}}.
%%CITATION = 1103.1887;%%.

\bibitem{Szendroi:2007nu}
B.~Szendroi, ``{Non-commutative Donaldson-Thomas theory and the conifold},''
  \href{http://dx.doi.org/10.2140/gt.2008.12.1171}{{\em Geom. Topol.}
  {\bfseries 12} (2008) 1171--1202},
\href{http://arxiv.org/abs/0705.3419}{{\ttfamily arXiv:0705.3419 [math.AG]}}.
%%CITATION = 0705.3419;%%.

\bibitem{Nagao:2010kx}
K.~Nagao and H.~Nakajima, ``{Counting invariant of perverse coherent sheaves
  and its wall-crossing},''
\href{http://arxiv.org/abs/0809.2992}{{\ttfamily arXiv:0809.2992 [math.AG]}}.
%%CITATION = 0809.2992;%%.

\bibitem{Nagao}
K.~Nagao, ``{Derived categories of small toric Calabi-Yau 3-folds and counting
  invariants},'' \href{http://arxiv.org/abs/0809.2994}{{\ttfamily
  arXiv:0809.2994 [math.AG]}}.

\bibitem{Collinucci:2008ht}
A.~Collinucci and T.~Wyder, ``{The elliptic genus from split flows and
  Donaldson-Thomas invariants},''
  \href{http://dx.doi.org/10.1007/JHEP05(2010)081}{{\em JHEP} {\bfseries 05}
  (2010) 081},
\href{http://arxiv.org/abs/0810.4301}{{\ttfamily arXiv:0810.4301 [hep-th]}}.
%%CITATION = 0810.4301;%%.

\bibitem{Jafferis:2008uf}
D.~L. Jafferis and G.~W. Moore, ``{Wall crossing in local Calabi Yau
  manifolds},''
\href{http://arxiv.org/abs/0810.4909}{{\ttfamily arXiv:0810.4909 [hep-th]}}.
%%CITATION = 0810.4909;%%.

\bibitem{Chuang:2008aw}
W.-y. Chuang and D.~L. Jafferis, ``{Wall Crossing of BPS States on the Conifold
  from Seiberg Duality and Pyramid Partitions},''
  \href{http://dx.doi.org/10.1007/s00220-009-0832-2}{{\em Commun. Math. Phys.}
  {\bfseries 292} (2009) 285--301},
\href{http://arxiv.org/abs/0810.5072}{{\ttfamily arXiv:0810.5072 [hep-th]}}.
%%CITATION = 0810.5072;%%.

\bibitem{Kontsevich-Soibelman}
M.~Kontsevich and Y.~Soibelman, ``{Stability structures, motivic
  Donaldson-Thomas invariants and cluster transformations},''
  \href{http://arxiv.org/abs/0811.2435}{{\ttfamily arXiv:0811.2435 [math.AG]}}.

\bibitem{Ooguri:2008yb}
H.~Ooguri and M.~Yamazaki, ``{Crystal Melting and Toric Calabi-Yau
  Manifolds},'' \href{http://dx.doi.org/10.1007/s00220-009-0836-y}{{\em Commun.
  Math. Phys.} {\bfseries 292} (2009) 179--199},
\href{http://arxiv.org/abs/0811.2801}{{\ttfamily arXiv:0811.2801 [hep-th]}}.
%%CITATION = 0811.2801;%%.

\bibitem{Dimofte:2009bv}
T.~Dimofte and S.~Gukov, ``{Refined, Motivic, and Quantum},''
  \href{http://dx.doi.org/10.1007/s11005-009-0357-9}{{\em Lett. Math. Phys.}
  {\bfseries 91} (2010) 1},
\href{http://arxiv.org/abs/0904.1420}{{\ttfamily arXiv:0904.1420 [hep-th]}}.
%%CITATION = 0904.1420;%%.

\bibitem{Chuang:2009pd}
W.-y. Chuang and G.~Pan, ``{BPS State Counting in Local Obstructed Curves from
  Quiver Theory and Seiberg Duality},'' {\em J. Math. Phys.} {\bfseries 51}
  (2010) 052305,
\href{http://arxiv.org/abs/0908.0360}{{\ttfamily arXiv:0908.0360 [hep-th]}}.
%%CITATION = 0908.0360;%%.

\bibitem{Kontsevich:2009xt}
M.~Kontsevich and Y.~Soibelman, ``{Motivic Donaldson-Thomas invariants: summary
  of results},''
\href{http://arxiv.org/abs/0910.4315}{{\ttfamily arXiv:0910.4315 [math.AG]}}.
%%CITATION = 0910.4315;%%.

\bibitem{Sulkowski:2009rw}
P.~Sulkowski, ``{Wall-crossing, free fermions and crystal melting},''
\href{http://arxiv.org/abs/0910.5485}{{\ttfamily arXiv:0910.5485 [hep-th]}}.
%%CITATION = 0910.5485;%%.

\bibitem{Dimofte:2009tm}
T.~Dimofte, S.~Gukov, and Y.~Soibelman, ``{Quantum Wall Crossing in N=2 Gauge
  Theories},'' \href{http://dx.doi.org/10.1007/s11005-010-0437-x}{{\em Lett.
  Math. Phys.} {\bfseries 95} (2011) 1--25},
\href{http://arxiv.org/abs/0912.1346}{{\ttfamily arXiv:0912.1346 [hep-th]}}.
%%CITATION = 0912.1346;%%.

\bibitem{Krefl:2010sz}
D.~Krefl, ``{Wall Crossing Phenomenology of Orientifolds},''
\href{http://arxiv.org/abs/1001.5031}{{\ttfamily arXiv:1001.5031 [hep-th]}}.
%%CITATION = 1001.5031;%%.

\bibitem{Aganagic:2010qr}
M.~Aganagic and K.~Schaeffer, ``{Wall Crossing, Quivers and Crystals},''
\href{http://arxiv.org/abs/1006.2113}{{\ttfamily arXiv:1006.2113 [hep-th]}}.
%%CITATION = 1006.2113;%%.

\bibitem{Collinucci:2009nv}
A.~Collinucci, P.~Soler, and A.~M. Uranga, ``{Non-perturbative effects and
  wall-crossing from topological strings},''
  \href{http://dx.doi.org/10.1088/1126-6708/2009/11/025}{{\em JHEP} {\bfseries
  11} (2009) 025},
\href{http://arxiv.org/abs/0904.1133}{{\ttfamily arXiv:0904.1133 [hep-th]}}.
%%CITATION = 0904.1133;%%.

\bibitem{Nagao:2009rq}
K.~Nagao and M.~Yamazaki, ``{The Non-commutative Topological Vertex and Wall
  Crossing Phenomena},''
\href{http://arxiv.org/abs/0910.5479}{{\ttfamily arXiv:0910.5479 [hep-th]}}.
%%CITATION = 0910.5479;%%.

\bibitem{Cecotti:2009uf}
S.~Cecotti and C.~Vafa, ``{BPS Wall Crossing and Topological Strings},''
\href{http://arxiv.org/abs/0910.2615}{{\ttfamily arXiv:0910.2615 [hep-th]}}.
%%CITATION = 0910.2615;%%.

\bibitem{Szabo:2009vw}
R.~J. Szabo, ``{Instantons, Topological Strings and Enumerative Geometry},''
  \href{http://dx.doi.org/10.1155/2010/107857}{{\em Adv. Math. Phys.}
  {\bfseries 2010} (2010) 107857},
\href{http://arxiv.org/abs/0912.1509}{{\ttfamily arXiv:0912.1509 [hep-th]}}.
%%CITATION = 0912.1509;%%.

\bibitem{Chuang:2010wx}
W.-y. Chuang, D.-E. Diaconescu, and G.~Pan, ``{Rank Two ADHM Invariants and
  Wallcrossing},''
\href{http://arxiv.org/abs/1002.0579}{{\ttfamily arXiv:1002.0579 [math.AG]}}.
%%CITATION = 1002.0579;%%.

\bibitem{Bringmann:2010sd}
K.~Bringmann and J.~Manschot, ``{From sheaves on P2 to a generalization of the
  Rademacher expansion},''
\href{http://arxiv.org/abs/1006.0915}{{\ttfamily arXiv:1006.0915 [math.NT]}}.
%%CITATION = 1006.0915;%%.

\bibitem{Manschot:2010nc}
J.~Manschot, ``{The Betti numbers of the moduli space of stable sheaves of rank
  3 on P2},''
\href{http://arxiv.org/abs/1009.1775}{{\ttfamily arXiv:1009.1775 [math-ph]}}.
%%CITATION = 1009.1775;%%.

\bibitem{Aganagic:2009kf}
M.~Aganagic, H.~Ooguri, C.~Vafa, and M.~Yamazaki, ``{Wall Crossing and
  M-theory},''
\href{http://arxiv.org/abs/0908.1194}{{\ttfamily arXiv:0908.1194 [hep-th]}}.
%%CITATION = 0908.1194;%%.

\bibitem{Aganagic:2009cg}
M.~Aganagic and M.~Yamazaki, ``{Open BPS Wall Crossing and M-theory},''
  \href{http://dx.doi.org/10.1016/j.nuclphysb.2010.03.019}{{\em Nucl. Phys.}
  {\bfseries B834} (2010) 258--272},
\href{http://arxiv.org/abs/0911.5342}{{\ttfamily arXiv:0911.5342 [hep-th]}}.
%%CITATION = 0911.5342;%%.

\bibitem{Cheng:2007ch}
M.~C.~N. Cheng and E.~Verlinde, ``{Dying Dyons Don't Count},''
  \href{http://dx.doi.org/10.1088/1126-6708/2007/09/070}{{\em JHEP} {\bfseries
  09} (2007) 070},
\href{http://arxiv.org/abs/0706.2363}{{\ttfamily arXiv:0706.2363 [hep-th]}}.
%%CITATION = 0706.2363;%%.

\bibitem{Sen:2008ht}
A.~Sen, ``{Wall Crossing Formula for N=4 Dyons: A Macroscopic Derivation},''
  \href{http://dx.doi.org/10.1088/1126-6708/2008/07/078}{{\em JHEP} {\bfseries
  07} (2008) 078},
\href{http://arxiv.org/abs/0803.3857}{{\ttfamily arXiv:0803.3857 [hep-th]}}.
%%CITATION = 0803.3857;%%.

\bibitem{Cheng:2008fc}
M.~C.~N. Cheng and E.~P. Verlinde, ``{Wall Crossing, Discrete Attractor Flow,
  and Borcherds Algebra},''
  \href{http://dx.doi.org/10.3842/SIGMA.2008.068}{{\em SIGMA} {\bfseries 4}
  (2008) 068},
\href{http://arxiv.org/abs/0806.2337}{{\ttfamily arXiv:0806.2337 [hep-th]}}.
%%CITATION = 0806.2337;%%.

\bibitem{Cheng:2008kt}
M.~C.~N. Cheng and A.~Dabholkar, ``{Borcherds-Kac-Moody Symmetry of N=4
  Dyons},'' {\em Commun. Num. Theor. Phys.} {\bfseries 3} (2009) 59--110,
\href{http://arxiv.org/abs/0809.4258}{{\ttfamily arXiv:0809.4258 [hep-th]}}.
%%CITATION = 0809.4258;%%.

\bibitem{Cheng:2009hm}
M.~C.~N. Cheng and L.~Hollands, ``{A Geometric Derivation of the Dyon
  Wall-Crossing Group},''
  \href{http://dx.doi.org/10.1088/1126-6708/2009/04/067}{{\em JHEP} {\bfseries
  04} (2009) 067},
\href{http://arxiv.org/abs/0901.1758}{{\ttfamily arXiv:0901.1758 [hep-th]}}.
%%CITATION = 0901.1758;%%.

\bibitem{Gaiotto:2008cd}
D.~Gaiotto, G.~W. Moore, and A.~Neitzke, ``{Four-dimensional wall-crossing via
  three-dimensional field theory},''
  \href{http://dx.doi.org/10.1007/s00220-010-1071-2}{{\em Commun. Math. Phys.}
  {\bfseries 299} (2010) 163--224},
\href{http://arxiv.org/abs/0807.4723}{{\ttfamily arXiv:0807.4723 [hep-th]}}.
%%CITATION = 0807.4723;%%.

\bibitem{Gaiotto:2009hg}
D.~Gaiotto, G.~W. Moore, and A.~Neitzke, ``{Wall-crossing, Hitchin Systems, and
  the WKB Approximation},''
\href{http://arxiv.org/abs/0907.3987}{{\ttfamily arXiv:0907.3987 [hep-th]}}.
%%CITATION = 0907.3987;%%.

\bibitem{Gaiotto:2010be}
D.~Gaiotto, G.~W. Moore, and A.~Neitzke, ``{Framed BPS States},''
\href{http://arxiv.org/abs/1006.0146}{{\ttfamily arXiv:1006.0146 [hep-th]}}.
%%CITATION = 1006.0146;%%.

\bibitem{Gaiotto:2011tf}
D.~Gaiotto, G.~W. Moore, and A.~Neitzke, ``{Wall-Crossing in Coupled 2d-4d
  Systems},''
\href{http://arxiv.org/abs/1103.2598}{{\ttfamily arXiv:1103.2598 [hep-th]}}.
%%CITATION = 1103.2598;%%.

\bibitem{Diaconescu:2007bf}
E.~Diaconescu and G.~W. Moore, ``{Crossing the Wall: Branes vs. Bundles},''
\href{http://arxiv.org/abs/0706.3193}{{\ttfamily arXiv:0706.3193 [hep-th]}}.
%%CITATION = 0706.3193;%%.

\bibitem{Jafferis:2007ti}
D.~L. Jafferis and N.~Saulina, ``{Fragmenting D4 branes and coupled q-deformed
  Yang Mills},''
\href{http://arxiv.org/abs/0710.0648}{{\ttfamily arXiv:0710.0648 [hep-th]}}.
%%CITATION = 0710.0648;%%.

\bibitem{Andriyash:2008it}
E.~Andriyash and G.~W. Moore, ``{Ample D4-D2-D0 Decay},''
\href{http://arxiv.org/abs/0806.4960}{{\ttfamily arXiv:0806.4960 [hep-th]}}.
%%CITATION = 0806.4960;%%.

\bibitem{David:2009ru}
J.~R. David, ``{On walls of marginal stability in N=2 string theories},''
  \href{http://dx.doi.org/10.1088/1126-6708/2009/08/054}{{\em JHEP} {\bfseries
  08} (2009) 054},
\href{http://arxiv.org/abs/0905.4115}{{\ttfamily arXiv:0905.4115 [hep-th]}}.
%%CITATION = 0905.4115;%%.

\bibitem{Manschot:2009ia}
J.~Manschot, ``{Stability and duality in N=2 supergravity},''
  \href{http://dx.doi.org/10.1007/s00220-010-1104-x}{{\em Commun. Math. Phys.}
  {\bfseries 299} (2010) 651--676},
\href{http://arxiv.org/abs/0906.1767}{{\ttfamily arXiv:0906.1767 [hep-th]}}.
%%CITATION = 0906.1767;%%.

\bibitem{Manschot:2010xp}
J.~Manschot, ``{Wall-crossing of D4-branes using flow trees},''
\href{http://arxiv.org/abs/1003.1570}{{\ttfamily arXiv:1003.1570 [hep-th]}}.
%%CITATION = 1003.1570;%%.

\bibitem{Chuang:2010ii}
W.-y. Chuang, D.-E. Diaconescu, and G.~Pan, ``{Wallcrossing and Cohomology of
  The Moduli Space of Hitchin Pairs},''
\href{http://arxiv.org/abs/1004.4195}{{\ttfamily arXiv:1004.4195 [math.AG]}}.
%%CITATION = 1004.4195;%%.

\bibitem{Nishinaka:2010qk}
T.~Nishinaka and S.~Yamaguchi, ``{Wall-crossing of D4-D2-D0 and flop of the
  conifold},'' \href{http://dx.doi.org/10.1007/JHEP09(2010)026}{{\em JHEP}
  {\bfseries 09} (2010) 026}, \href{http://arxiv.org/abs/1007.2731}{{\ttfamily
  arXiv:1007.2731 [hep-th]}}.

\bibitem{Nishinaka:2010fh}
T.~Nishinaka, ``{Multiple D4-D2-D0 on the Conifold and Wall-crossing with the
  Flop},'' \href{http://arxiv.org/abs/1010.6002}{{\ttfamily arXiv:1010.6002
  [hep-th]}}.

\bibitem{Alim:2010cf}
M.~Alim, B.~Haghighat, M.~Hecht, A.~Klemm, M.~Rauch, and T.~Wotschke,
  ``{Wall-crossing holomorphic anomaly and mock modularity of multiple
  M5-branes},''
\href{http://arxiv.org/abs/1012.1608}{{\ttfamily arXiv:1012.1608 [hep-th]}}.
%%CITATION = 1012.1608;%%.

\bibitem{Nishinaka:2011sv}
T.~Nishinaka and S.~Yamaguchi, ``{Statistical model and BPS D4-D2-D0
  counting},'' \href{http://arxiv.org/abs/1102.2992}{{\ttfamily arXiv:1102.2992
  [hep-th]}}.

\bibitem{Uranga:1998vf}
A.~M. Uranga, ``{Brane Configurations for Branes at Conifolds},'' {\em JHEP}
  {\bfseries 01} (1999) 022,
\href{http://arxiv.org/abs/hep-th/9811004}{{\ttfamily arXiv:hep-th/9811004}}.
%%CITATION = HEP-TH/9811004;%%.

\bibitem{Bershadsky:1995sp}
M.~Bershadsky, C.~Vafa, and V.~Sadov, ``{D-Strings on D-Manifolds},''
  \href{http://dx.doi.org/10.1016/0550-3213(96)00024-7}{{\em Nucl. Phys.}
  {\bfseries B463} (1996) 398--414},
\href{http://arxiv.org/abs/hep-th/9510225}{{\ttfamily arXiv:hep-th/9510225}}.
%%CITATION = HEP-TH/9510225;%%.

\end{thebibliography}
\end{document}